\documentclass[aps,prl,superscriptaddress,floatfix,reprint]{revtex4-1}
\usepackage{amssymb}
\usepackage{amsmath}
\usepackage{graphicx}
\usepackage{color}
\usepackage{siunitx}
\usepackage{hyperref}

\begin{document}

\title{\Large Quantum dephasing in a gated GaAs triple quantum dot due to nonergodic noise
}

\author{M.R.~Delbecq}
\email[email: ]{matthieu.delbecq@riken.jp}
\affiliation{Center for Emergent Matter Science, RIKEN, 2-1 Hirosawa, Wako-shi, Saitama, 351-0198, Japan}

\author{T.~Nakajima}
\affiliation{Center for Emergent Matter Science, RIKEN, 2-1 Hirosawa, Wako-shi, Saitama, 351-0198, Japan}

\author{P.~Stano}
\affiliation{Center for Emergent Matter Science, RIKEN, 2-1 Hirosawa, Wako-shi, Saitama, 351-0198, Japan}
\affiliation{Institute of Physics, Slovak Academy of Sciences, 845 11 Bratislava, Slovakia}

\author{T.~Otsuka}
\affiliation{Center for Emergent Matter Science, RIKEN, 2-1 Hirosawa, Wako-shi, Saitama, 351-0198, Japan}

\author{S.~Amaha}
\affiliation{Center for Emergent Matter Science, RIKEN, 2-1 Hirosawa, Wako-shi, Saitama, 351-0198, Japan}

\author{J.~Yoneda}
\affiliation{Center for Emergent Matter Science, RIKEN, 2-1 Hirosawa, Wako-shi, Saitama, 351-0198, Japan}

\author{K.~Takeda}
\affiliation{Center for Emergent Matter Science, RIKEN, 2-1 Hirosawa, Wako-shi, Saitama, 351-0198, Japan}

\author{G.~Allison}
\affiliation{Center for Emergent Matter Science, RIKEN, 2-1 Hirosawa, Wako-shi, Saitama, 351-0198, Japan}

\author{A.~Ludwig}
\affiliation{Lehrstuhl f\"{u}r Angewandte Festk\"{o}rperphysik, Ruhr-Universit\"{a}t Bochum, D-44780 Bochum, Germany}

\author{A.D.~Wieck}
\affiliation{Lehrstuhl f\"{u}r Angewandte Festk\"{o}rperphysik, Ruhr-Universit\"{a}t Bochum, D-44780 Bochum, Germany}

\author{S.~Tarucha}
\affiliation{Center for Emergent Matter Science, RIKEN, 2-1 Hirosawa, Wako-shi, Saitama, 351-0198, Japan}
\affiliation{Department of Applied Physics, University of Tokyo, 7-3-1 Hongo, Bunkyo-ku, Tokyo, 113-8656, Japan}

\date{\today}

\begin{abstract}
We extract the phase coherence of a qubit defined by singlet and triplet electronic states in a gated GaAs triple quantum dot, measuring on timescales much shorter than the decorrelation time of the environmental noise. In this non-ergodic regime, we observe that the coherence is boosted and several dephasing times emerge, depending on how the phase stability is extracted. We elucidate their mutual relations, and demonstrate that they reflect the noise short-time dynamics.
\end{abstract}

\maketitle

Noise induces dephasing and loss of coherence of quantum systems. The finite resonance signal linewidth in the nuclear magnetic resonance (NMR) \cite{Abragam} or electron spin resonance (ESR) \cite{Poole1993} experiment is one of its paradigmatic manifestations, allowing one to infer the corresponding dephasing time $T_2^\star$. As such experiments are usually performed in the steady state and on large ensembles of spins, this dephasing time reflects the system inhomogeneity over a large range both in space and time.

This dephasing is a central issue for further progress of quantum information science \cite{DiVincenzo1995}. In electronic spin qubits realized in semiconductor quantum dots \cite{Loss1998,Petta2005}, the dominant noise is often the thermally fluctuating Overhauser field of nuclear spins \cite{Merkulov2002,Khaetskii2002}. The hall-mark of this environment is its very slow internal dynamics \cite{Urbaszek2013}, due to the weakness of nuclear spin-spin interactions \cite{Paget1977,Maletinsky2007a}. This slowness allows strong suppression of the arising qubit dephasing by dynamical decoupling \cite{Bluhm2010,DeSousa2009}, or Hamiltonian estimation \cite{Shulman2014}, techniques based on the ability to operate the qubit on times much shorter than the noise decorrelation time. This is a very different regime than that of the steady state NMR/ESR measurements, and one expects that the extracted dephasing might be strongly affected. We exploit the solid state qubit technology with its fast and sensitive readout techniques, to access dephasing in this regime. We investigate the nature of $T_2^\star$, which becomes a dynamical quantity itself, and its relation to the underlying noise dynamics.

We probe the coupled electron-nuclei system on timescales well below the nuclear spins decorrelation time, building on methods developed in Ref.~\onlinecite{Shulman2014}. Concerning nuclei, we find a striking sub-diffusive behaviour of the Overhauser field correlator, at odds with existing theories. Concerning the qubit, we demonstrate that the dephasing time depends sensitively on the way the coherence is measured. While in the ergodic regime the variance $\sigma^2_B$ of the Overhauser field $B_N$ gives the qubit dephasing time as $1/(\pi \sqrt{2} \sigma_B)$, we find a larger phase coherence in the non-ergodic regime. In addition, the phase coherence becomes a stochastic variable with a non-trivial probability distribution. Finally, working in the non-ergodic regime of a diffusive noise, a tenfold decrease in the measurement time automatically prolongs the qubit phase coherence by roughly a factor of 3.

\begin{figure}[!]
\includegraphics[width=0.48\textwidth]{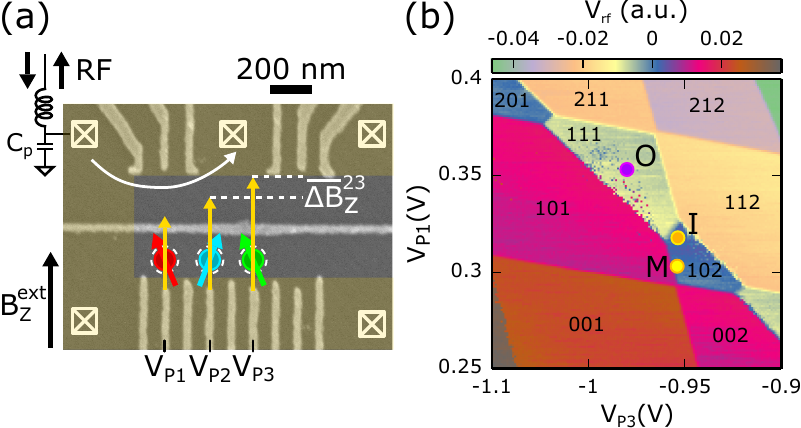}
\caption{(color online). (a) SEM micrograph of a similar device to the one measured. Lateral gates defining quantum dots (bottom) and charge sensors (top) are shown in light grey on the dark grey surface of the GaAs substrate. The three leftmost quantum dots are formed and manipulated while the upper left charge sensor, connected to an rf-reflectometry circuit is used. The ``C-shaped'' light colored area denotes the micromagnet providing inhomogenous magnetic field. An external magnetic field $B_{z}^{ext}=0.7$ T is applied. (b) Charge stability diagram in the plane defined by plunger gates $\mathrm{P_1}$ and $\mathrm{P_3}$. The positions for initialization (I), operation (O) and measurement (M) configurations are denoted. } \label{Fig:1}
\end{figure}

Our device is a triple spin qubit shown in Fig.~\ref{Fig:1}(a). A micromagnet generates a magnetic field difference $\Delta B_{MM}^{z}$ between the dots \cite{Pioro-Ladriere2008}. Working between the (1,0,2) and (1,1,1) charge configurations, we manipulate the two rightmost dots as a singlet-triplet qubit \cite{Petta2005,Hanson2007} and leave the leftmost spin qubit idle [see Fig.~\ref{Fig:1}(b)]. The oscillation frequency of the singlet-triplet qubit $f=|g|\mu_B \Delta B_z/2 \pi \hbar$ (throughout the article we convert $\Delta B_z$ to frequency with this formula using g-factor $g=-0.44$) is set by the magnetic field gradient $\Delta B_z = \Delta B_{MM}^{z} + \Delta B_{nuc}^z$, thus subject to nuclear field fluctuations. 

We extract the qubit dephasing time $T_2^\star$ from the free induction decay, organizing the measurement scheme into the following hierarchy. The basic unit is a ``cycle'' (index $c$) during which the qubit is initialized in the state $\left|\uparrow,\mathrm{S(0,2)}\right\rangle$, then quickly moved to the $\left|\uparrow,\mathrm{S(1,1)}\right\rangle$ state where it precesses with $\left|\uparrow,\mathrm{T_0(1,1)}\right\rangle$ for the qubit evolution time $\tau_c$ before undergoing a Pauli spin blockade measurement deep in the (1,0,2) region \cite{Barthel2012}. The cycle duration is set to $\SI{15.192}{\micro\second}$ independent of $\tau_c$ by adjusting the initialization time. The next level is a ``record'', which comprises $250$ consecutive cycles with qubit evolution times increased by $\SI{4}{\nano\second}$ steps, restarting each record from zero. A single record takes time $t_{rec}=\SI{3.8}{\milli\second} = 250 \times \SI{15.192}{\micro\second}$ to acquire, covering the qubit evolution for $\tau_c \in [0,996]$ ns. Finally, we form a set $\mathcal{R}$ by selecting $N_\mathcal{R}$ records from all measured ones. We extract the projection of the qubit state on the $S$-$T_0$ axis of the Bloch sphere, $s(\tau_c)$, by averaging over data in $\mathcal{R}$, using $s(\tau_c)=\langle 2 P_S(\tau_{c}) - 1 \rangle_{\mathcal{R}}$ with $P_S\in\{ 0,1\}$ the results of projective measurements of the singlet state. The simplest choice is to take $\mathcal{R}$ as a block of $N$ consecutive records. The time to acquire such data is $\Delta t = N t_{rec}$,  referred to as the {\it acquisition time} in further. We select  $\mathcal{R}$ also in other ways below, but it always contains such blocks of $N$ consecutive records. It defines the acquisition time $\Delta t$ as a natural parameter for dephasing rates.

Indeed, even though we are interested in the qubit evolution on times of the order of $T_2^\star$, the acquisition time needed to sample the continuous function $s(\tau_c)$ from binary data of projective measurements is typically orders of magnitude larger, as is clear from the above measurement description. Now, if the acquisition time $\Delta t$ is so large that the values of the fluctuating Overhauser field $B_N(t)$ and $B_N(t + \Delta t)$ are uncorrelated, the measurement is in the ergodic regime and always yields the same dephasing time $T_{2,\infty}^\star$ \cite{Petta2005,Merkulov2002,Khaetskii2002}. Our measurement is in the opposite -- non-ergodic -- regime, with the noise decorrelation time much larger than the acquisition time. Here one generally expects longer coherence \cite{Shulman2014} and non-trivial signatures of the noise dynamics reflected in the obtained $T_2^\star$ value. 

\begin{figure}[!]
\includegraphics[width=0.48\textwidth]{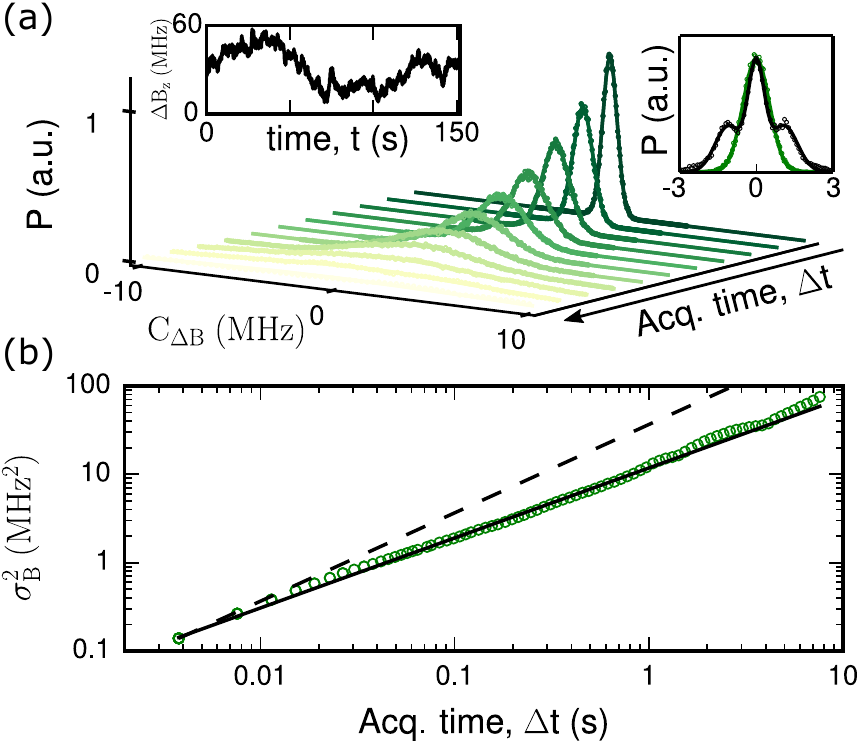}
\caption{(color online). (a) The probability distribution of the nuclear field gradient time correlator $C_{\Delta B}(\Delta t)$ for acquisition time $\Delta t$ from 3.8 ms (dark green) to 7.6 s (yellow). Data (dots) are fitted with a Gaussian distribution (line). Left inset: Nuclear field gradient $\Delta B_z(t)$ extracted from the qubit frequency as a function of time. Right inset: correlator for $\Delta t=\SI{3.8}{\milli\second}$ excluding (not excluding) the third spin fluctuation in green (black). (b) Variance of the nuclear field gradient correlator as a function of the acquisition time $\Delta t$. The solid line is a fit showing a growth with a power law exponent $\alpha = 0.8$. The dashed line shows a power law behavior with $\alpha = 1$ for comparison. } \label{Fig:2}
\end{figure}

We first extract the time evolution of $\Delta B_{z}$ over 40000 consecutive records spanning more than $\SI{2}{\minute}$ [Fig.~\ref{Fig:2}(a)]. It fluctuates around a finite value of $\SI{30}{\mega\hertz}$, set by the micromagnet, by $\pm \SI{20}{\mega\hertz}$ due to nuclei. With our measurement sequence we can follow the nuclei dynamics down to the time $t_{rec}$. Namely, using a Bayesian estimation algorithm \cite{Sergeevich2011, Shulman2014} on the data of a single record, we estimate the mean and variance of the qubit frequency as it evolved during that record. The correlator $C_{\Delta B}(\Delta t)=\Delta B_{nuc}^z(t+\Delta t) - \Delta B_{nuc}^z(t)$, shown in Fig.~\ref{Fig:2}(a), displays a clear Gaussian probability distribution which broadens as the acquisition time $\Delta t$ increases. As shown in Fig.~\ref{Fig:2}(b), its variance grows as $\sigma^2_{B}(\Delta t)=D (\Delta t)^\alpha$ over more than three orders of magnitude of timespan, with $\alpha = 0.8$ and $D=\SI[per-mode=symbol]{0.048}{\mega\hertz\tothe{2}\per\milli\second\tothe{0.8}}$. Though we do not reach such long times in our measurement, the growth has to saturate, at $\sigma^2_{B}(\infty)$, since the fluctuating Overhauser field is bounded. Taking a value $\sigma_{B}(\infty)$ corresponding to $T_2^\star = \SI{10}{\nano\second}$ typical for dots comparable to ours \cite{Petta2005}, we can roughly estimate the nuclear decorrelation time as $(\sigma_B^2(\infty)/D)^{1/\alpha} \approx  \SI{107}{\second}$. For GaAs, values from seconds to hours are reported, the large range being due to effects of doping, strain, and nanostructure confinement \cite{Urbaszek2013}. 

More interestingly, the exponent $\alpha<1$ indicates a surprising sub-diffusive behavior. This differs from the normal diffusion (corresponding to $\alpha=1$ \cite{Wang1945}) that is assumed \cite{Redfield1959} for dipole-dipole interactions that should dominate at times equal or larger than our $t_{rec}$, and super-diffusion expected for electron-mediated interactions which should dominate at much shorter times \cite{Klauser2008}. Non-Markovian nuclear dynamics could result in such sub-diffusion \cite{Bouchaud1990}, it would however also imply a non-Gaussian noise correlator \cite{Metzler2000}, at odds with our observations. Since it is difficult to infer the correlator functional form in the time domain from its noise power spectrum \cite{Li2012} if the latter is known only within a limited frequency range, previous investigations \cite{Reilly2008,Medford2012} do not necessarily contradict our observation. We also cannot completely exclude, due to our limited resolution, a behavior closer to standard diffusion at the shortest times we reach (Ref.~\onlinecite{Shulman2014} reports $\alpha=1$ for times below $\SI{50}{ms}$).

We now turn to the qubit phase stability. The standard way is to fit the qubit evolution to oscillations with a Gaussian decay
\begin{equation}
s(\tau_c)\stackrel{\rm fit}{\longrightarrow} \cos(2\pi f_0 \tau_c) \exp \left[ -\left( \frac{\tau_c}{T_2^\star} \right)^2 \right],
\label{eq:fit}
\end{equation}
and define the dephasing time as the fitted decay parameter. If $\tau_c$ is much smaller than the acquisition time, always fulfilled here, the frequency change during the time $\tau_c$ is negligible and we get
\begin{equation} 
s(\tau_c) = \frac{1}{N_\mathcal{R}} \sum_{r \in \mathcal{R}} \cos\left(2\pi f_{c,r} \tau_c\right),
\label{eq:set}
\end{equation}
with $f_{c,r}$ the qubit frequency during the $c$-th cycle of the $r$-th record. From here it follows that the frequency and dephasing extracted from the fit in Eq.~\eqref{eq:fit} are given, respectively, as the average and the variance of the set of frequencies $\{f_{c,r}\}$. These statistical properties in turn depend on how the set $\mathcal{R}$ is chosen.

\begin{figure}[!]
\includegraphics[width=0.48\textwidth]{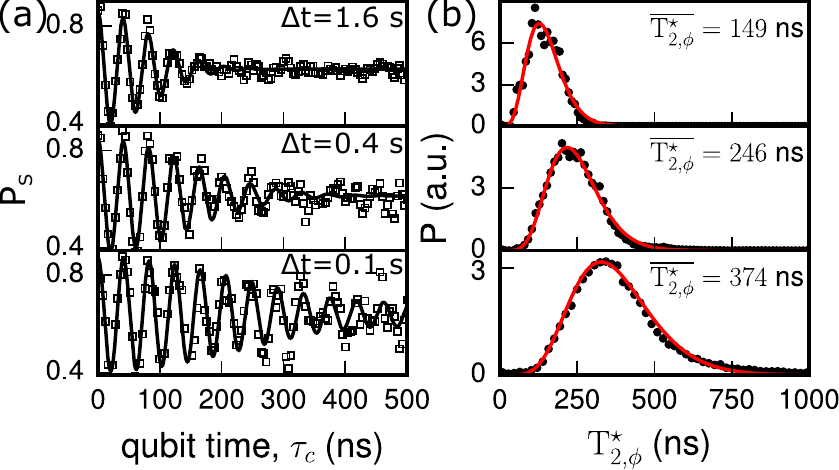}
\caption{(color online). (a) Typical qubit evolution traces for different acquisition times. Solid lines are fit to decaying oscillations giving $T_{2,\phi}^\star=$ 120, 220 and 570 ns respectively. (b) Probability density distributions of $T_{2,\phi}^\star$ corresponding to the same acquisition times as for (a). The red solid line is a fit to a Gamma distribution resulting in skewness $\gamma_1\approx0.75$, and $\overline{T_{2,\phi}^\star}\approx$ as given.} \label{Fig:3}
\end{figure}

The standard way is to choose $\mathcal{R}$ as a single block of $N$ consecutive records. Doing so we define $T^\star_{2,\phi}$, and observe a gradual increase of $T_{2,\phi}^\star\sim$ 120, 220 and 570 ns upon decreasing $N$, for acquisition times $\Delta t\sim$ 1.6, 0.4 and 0.1 s, see Fig.~\ref{Fig:3}(a). Since each of these qubit evolutions results from a particular noise realization, $T_2^\star$ becomes a stochastic variable itself. We are able to extract its probability distribution for various acquisition times, as shown in Fig.~\ref{Fig:3}(b). It is always well fitted by a Gamma distribution \cite{EPAPS} whose skewness does not significantly change for $\Delta t$ varying from $\SI{38}{\milli\second}$ to $\SI{7.6}{\second}$. We interpret this robustness as a signature that the nature of the underlying dynamics of nuclei does not change within this timespan. We conclude that a single trace is not sufficient to reliably estimate the phase decay, as the most probable $T_{2,\phi}^\star$ is smaller than the mean $\overline{T_{2,\phi}^\star}$, whereas occurrences of $T_{2,\phi}^\star$ several times larger than $\overline{T_{2,\phi}^\star}$ are common.

\begin{figure}[!]
\includegraphics[width=0.48\textwidth]{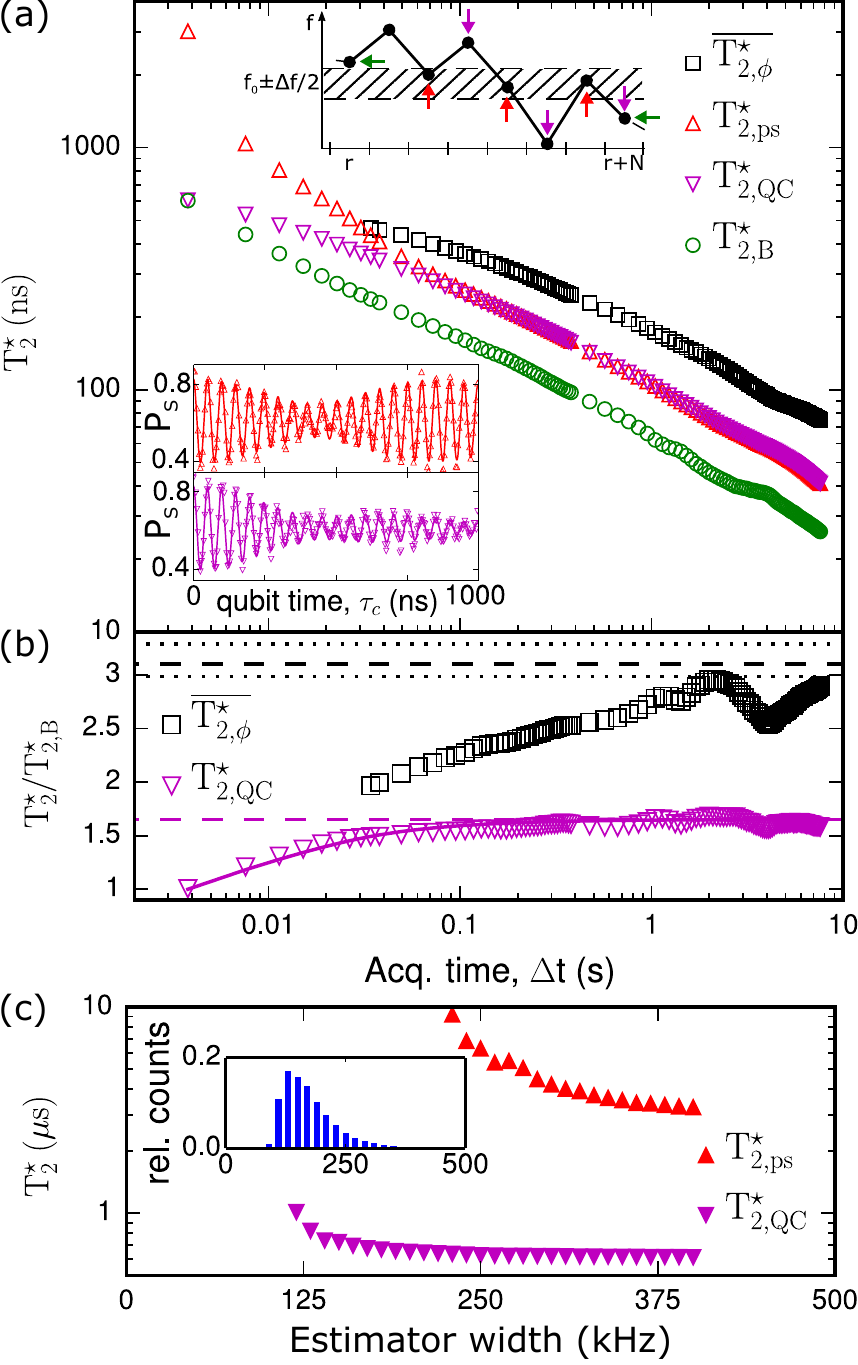}
\caption{(color online). (a) $T_2^\star$ from the different processings as a function of acquisition time $\Delta t$. Top inset: estimated frequencies of eight consecutive records. Averaging over all defines $T_{2,\phi}^\star$. The correlator $\sigma_B^2$ is calculated from the frequency difference between the first and last record (horizontal green arrow). Post-selection is performed by averaging over all blocks of records that start in a frequency window $f_0\pm \Delta f/2$ (upward red arrows), giving for $N=1$ the red trace in the lower inset defining $T_{2,ps}^\star$. (In this illustration, we neglect the complication of blocks overlaps that can happen for $N>1$). Blocks of records defining $T_{2,QC}^\star$ are those of $T_{2,ps}^\star$ shifted by 1 record (downward purple arrows), giving for $N=1$ the purple trace in the lower inset. (b) Ratios $\overline{T_{2,\phi}^\star}/T_{2,B}^\star$ (black squares) and $T_{2,QC}^\star/T_{2,B}^\star$ (purple triangles). Black dashed line (dotted black line): analytical limit for large $N$ for the mean shape ratio $k=7.25$ (for $k \in \langle 6, 8.5 \rangle$). Purple dashed line: analytical limit for large $N$. Purple solid line: numerical evaluation of the integral leading to Eq.~\eqref{eq:ratio2} \cite{EPAPS}. (c) Enhancement of $T_{2,ps}^\star$ (red upwards triangles) and $T_{2,QC}^\star$ (purple downwards triangles) by thresholding on the width of the estimator probability distribution peak. Inset: distribution of estimator probability peak width.} \label{Fig:4}
\end{figure}

This method is limited by the inherent noise of the quantum mechanical projective measurement and readout errors, the impacts of which increase as $N_\mathcal{R}$ decreases. We find that a minimum of ten records is required to get a reliable $\overline{T_{2,\phi}^\star}$. To access dephasing below this limit, we use a post-selection method, see the top inset of Fig.~\ref{Fig:4}(a). We include in $\mathcal{R}$ all blocks of $N$ consecutive records for which the Bayesian estimated frequency of the first one is within $f_0 \pm \Delta f/2$ (upward red arrows). For example, choosing $f_0=\SI{20}{\mega\hertz}$ and $\Delta f = \SI{0.1}{\mega\hertz}$ gives 167 such blocks, resulting for $N=1$ in the red trace shown in the lower inset of Fig.~\ref{Fig:4}(a), giving a coherence time $T_{2,ps}^\star \sim \SI{3}{\micro\second}$. Strikingly, we observe a beating pattern with frequency $\delta f \approx \SI{1.16}{\mega\hertz}$. Both the beating frequency and amplitude are consistent with thermal flips of the tunnel coupled spin in the leftmost dot, which lead to discrete jumps of the qubit oscillation frequency \cite{EPAPS}. The beating could only be unravelled thanks to the long coherence time we reach.

The dephasing times described above demonstrate a significant improvement compared to the 10 ns observed in the ergodic regime, but they cannot be taken as the measure of phase stability for general quantum computation (QC) algorithms. Indeed, the qubit oscillation frequency $f_0$ is only known after the fit in Eq.~\eqref{eq:fit} is performed, and therefore the measurement is finished, limiting its practical use for post-processing or echo techniques \cite{Hahn1950,Vandersypen2005,Hanson2007}. To access the dephasing time of a qubit whose frequency is known in advance \cite{Shulman2014}, which we denote as $T_{2,QC}^\star$, we select blocks by beginning with the records \emph{following} those with frequency $f_0\pm \Delta f$ [downward purple arrows in the inset of Fig.~\ref{Fig:4}(a)]. This set $\mathcal{R}$ can be thus obtained from the one in the previous paragraph by shifting all the records indexes by 1. The resulting trace is shown in purple in the lower inset of Fig.~\ref{Fig:4}(a) with $T_{2,QC}^\star \sim \SI{600}{\nano\second}$. The comparison of the different dephasing times is summarized on Fig.~\ref{Fig:4}(a) as a function of the acquisition time. We also include the nuclear field correlator variance through $T_{2,B}^\star(\Delta t)=1/[\pi\sqrt{2}\sigma_{B}(\Delta t)]$, the relation valid in the ergodic regime. 

The relations between these quantities are governed by the nuclear field dynamics. Approximating the nuclear dynamics as a random walk ($\alpha=1$), we were able to derive the following analytical results, valid for large $N$, hence long acquisition times, but still in the non-ergodic regime (see Ref.~\cite{EPAPS} for details). First,
\begin{equation}
\frac{\overline{T_{2,\phi}^\star}}{{T_{2,B}^\star}}=\sqrt{6} \, \frac{k}{\sqrt{(k-1)(k-2)}}  \approx 3,
\label{eq:ratio1}
\end{equation}
with $k=4/\gamma_1^2$ given by the skewness $\gamma_1$ of the Gamma distribution of $T_{2,\phi}^\star$ and we used $k=7.5$ to evaluate the ratio. The measured values are shown in Fig.~\ref{Fig:4}(b) as black squares while Eq.~\eqref{eq:ratio1} is shown by a black dashed line, showing the expected agreement for $N\gg 1$ with small deviations. Second, 
\begin{equation}
\frac{T_{2,QC}^\star}{T_{2,B}^\star} =  \sqrt{\frac{1}{2} \left(8 \sqrt{2} - 1 - \sqrt{1 + 16\sqrt{2}} \right)} \approx 1.65.
\label{eq:ratio2}
\end{equation}
The ratio extracted from the measurement is shown in Fig.~\ref{Fig:4}(b) as purple downward triangles. The $N\gg 1$ limit is displayed as a purple dashed line and a straightforward numerical calculation for finite $N$ as a purple solid line, showing excellent agreement with the data. 

We can further enhance the qubit coherence by constraining the selected records according to progressively smaller widths of the Bayesian estimator probability distribution. As shown in Fig.~\ref{Fig:4}(c), this boosts $T_{2,QC}^\star$ beyond $\SI{1}{\micro\second}$ by better estimating the oscillation frequency $f_0$. Even though similar or even larger values have been reported in GaAs \cite{Shulman2014} or other materials \cite{Maune2012,Kawakami2014,Veldhorst2014}, our architecture is explicitly a multi-qubit one. The presence of the third spin, which was probably the main limitation of the Bayesian estimator precision [see right inset of Fig.~\ref{Fig:2}(a)] \cite{EPAPS}, nevertheless manifestly proves that GaAs provides a robust platform for scalable architectures \cite{Delbecq2014,Otsuka2015a} with long coherence times. In addition, the qubit-qubit coupling we see offers resources for quantum computation, e.g. allowing implementation of entangling gates.

We would like to point out that one should be cautious about an apparent enhancement of the phase stability obtained by sophisticated post-processing. As an example, we can push $T_{2,ps}^\star$ up to $\SI{10}{\micro\second}$, using the post-selection described in the previous paragraph by which we effectively select records with especially low noise history. With little relevance for practical quantum computation, it nevertheless allows us to move towards the quantum mechanical limit set by $T_2$ which was argued to be much shorter for a free induction decay (our experiment) than in a Hahn echo sequence \cite{Yao2006} where $T_2\sim \SI{30}{\micro\second}$ has been reported \cite{Bluhm2010}. As we see no apparent saturation of $T_2^\star$ in the post-selection despite our sample not being optimized to maximize $T_2^\star$, we believe that the dephasing time will be further increased by straightforwardly reducing the acquisition time. This should allow access to both the quantum mechanical decay of the spin qubit and short-time dynamics of nuclei. Both are open problems with many interesting theoretical predictions which await experimental investigation \cite{Coish2004,Deng2006,Klauser2008}.

\begin{acknowledgments}
We thank T. Fujisawa, T. Kontos, D. Loss, Y. Nakamura, and Y. Tokura for helpful discussions and comments on the manuscript. This work  was supported financially by the ImPACT Program of Council for Science, Technology and Innovation, IARPA project ``Multi-Qubit Coherent Operations'' through Copenhagen University, CREST program of Japan Science and Technology Agency, JSPS Grant-in-Aid for, Scientific Research S (No. 26220710), Scientific Research B (No. 15H03524) and Research Young Scientists B (No. 25790006 \& No. 25800173), RIKEN Incentive Research Project, Toyota Physical \& Chemical Research Institute Scholars, Yazaki Memorial Foundation for Science and Technology Research Grant, Japan Prize Foundation Research Grant, Advanced Technology Institute Research Grant, Murata Science Foundation Research Grant and Izumi Science and Technology Foundation Research Grant. A. D. W. and A. L. gratefully acknowledge support from Mercur Pr2013-0001, BMBF Q.Com-H 16KIS0109, TRR160, and DFH/UFA CDFA-05-06. P.S. acknowledges support from APVV-0808-12(QIMABOS).
\end{acknowledgments}

\newpage
\begin{widetext}

\renewcommand{\figurename}{Figure S\!\!}
\setcounter{figure}{0}

\begin{center}
{\Large\bfseries Supplemental Material to\\ `Quantum dephasing in a gated GaAs triple quantum dot due to nonergodic noise'}
\end{center}

\section{EXPERIMENTAL DETAILS} \label{epaps:xp_details}
The experiments are performed at a base temperature of $\SI{20}{\milli\kelvin}$ in  a dilution refrigerator. The electronic temperature $T_{el}\sim \SI{300}{\milli\kelvin}$ has been obtained independently by transport measurements in the Coulomb blockade regime (data not shown). Voltage pulses applied on the gates of the TQD device are generated by a Tektronix 70002A Arbitrary Waveform Generator. The rf-QPC demodulated signal of the spin blockade measurement is digitized with an AlazarTech ATS9440 at a 125 MSample/s sampling rate. The qubit initialization time is $t_I=\SI{7.192}{\micro\second} - \tau_c$. The measurement time is $t_M = \SI{4}{\micro\second}$ to which we add a compensation pulse of $\SI{4}{\micro\second}$ so that the pulse sequence does not create a DC shift of the gate voltages, by setting the integral of the pulse sequence to 0. The compensation pulse length is adjusted so that the working point remains close to initialization and measurement points in the (1,0,2) region. The digitized measurement signal is integrated and thresholded to distinguish singlet from triplet states. Drifts in the charge sensor signal are accounted for by tracking the histograms of singlet and triplet states over $\SI{380}{\milli\second}$ corresponding to 100 records.

\section{ENERGY SPECTRUM OF OUR SYSTEM} \label{epaps:energy_spectrum}
The following Hamiltonian describes our system
\begin{eqnarray*}
H = & & \sum_{\alpha\beta} \left(t_{\alpha\beta} c_{\alpha\sigma}^\dagger c_{\beta\sigma} + t_{\alpha\beta}^* c_{\beta\sigma}^\dagger c_{\alpha\sigma} \right) \nonumber \\*
& + & \sum_i g\mu_B \mathbf{B}_i \cdot \mathbf{S}_i \nonumber \\*
& + & \sum_{\sigma=\uparrow,\downarrow} \frac{\epsilon - \epsilon_L}{2} \left|S(2,0),\sigma\right\rangle \left\langle S(2,0),\sigma\right| \nonumber \\*
& - & \sum_{\sigma=\uparrow,\downarrow} \frac{\epsilon - \epsilon_R}{2} \left|\sigma,S(0,2)\right\rangle \left\langle \sigma,S(0,2)\right|,
\end{eqnarray*}
where $c_{\alpha \sigma}$ ($c_{\alpha \sigma}^\dagger$) is the annihilation (creation) operator of an electron in dot $\alpha$ with spin $\sigma$, $t_{\alpha\beta}$ is the nearest neighbour tunnel coupling between dots $\alpha$ and $\beta$, $\mathbf{B}_i = \mathbf{B}_{ext}+\mathbf{B}_{MM,i} + \mathbf{B}_{N,i}$ is the total magnetic field in dot $i$ with $\mathbf{B}_{MM}$ the micromagnet field and $\mathbf{B}_{N}$ the Overhauser field of nuclear spins, and $\epsilon$ is the detuning between dots 1 and 3, with $\epsilon_L$ ($\epsilon_R$) its value at the (2,0,1)-(1,1,1) [(1,0,2)-(1,1,1)] charge transition. The Hilbert space is restricted to the eight different spin configurations of the (1,1,1) charge state plus $\left|\uparrow,\mathrm{S(0,2)}\right\rangle$, $\left|\downarrow,\mathrm{S(0,2)}\right\rangle$, $\left|\mathrm{S(2,0), \uparrow}\right\rangle$ and $\left|\mathrm{S(2,0), \downarrow}\right\rangle$. The diagram of Fig.~S\ref{Fig:S0} was calculated by diagonalizing $H$ with parameters chosen so as to emphasize the energy level configuration. Using $H$ with parameters realistic for our device, we estimate that the singlet-triplet energy difference converted to frequency $\delta f$ differs for the two branches $\left| \uparrow \right\rangle \otimes \left|ST\right\rangle$ and $\left| \downarrow \right\rangle \otimes \left|ST\right\rangle$ by a value of the order of MHz due to a three spin exchange interaction, as observed in our data (see section ``Extraction of $\Delta B_z$'' and Fig.~S\ref{Fig:S1}(a)).

\begin{figure}[!]
\includegraphics[width=0.25\textwidth]{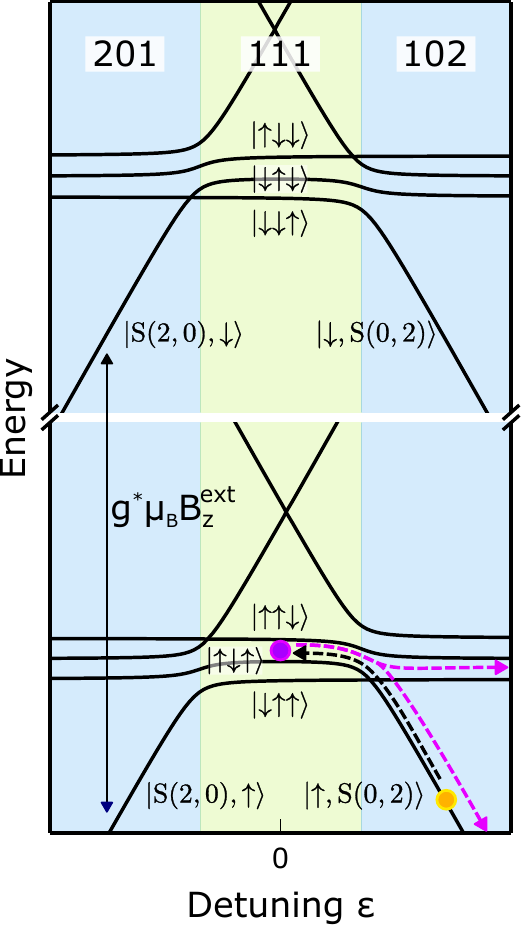}
\caption{Energy diagram of the triple spin qubit system under an inhomogenous Zeeman splitting for the three spins. The yellow circle indicates the initialization point of the qubit and the purple circle the operation point as in Fig.~1(b) of the main text. The dashed black arrow indicates the pulse from the initialization point to the operation point. The dashed purple arrow indicates the pulse towars the measurement points on either the singlet branch or on the triplet branch depending on the qubit state.} \label{Fig:S0}
\end{figure}

\section{EXTRACTION OF $\Delta B_z$} \label{epaps:extraction_delta_b}
Measurements are organized into blocks called records indexed by $r$, while the measurements within a single record are indexed by $r=1,\ldots,M$. A record comprises $M=250$ single-shot measurements of the qubit state after it evolved for time $\tau_{c,r}=\tau_c = 4(c-1)\ \si{\nano\second}$ with possible results $p_{c,r} = \pm 1$. We use the Bayesian formula \cite{Sergeevich2011,Shulman2014}
\begin{eqnarray} \label{eq:Bayes}
P&&(\Delta B_z^r|p_{M,r},p_{M-1,r},...,p_{1,r}) \nonumber \\*
&&\propto \prod_{r=1}^{M}\left[1+p_{c,r}(\alpha + \beta \cos(2\pi\Delta B_z^r \tau_c)) \right], 
\end{eqnarray}
to calculate the probability distribution $P$ for the value of the field gradient $\Delta B_z^r$ in record $r$. Here $\alpha$ and $\beta$ quantify errors due to imperfect initialization, the deviation of the qubit rotation axis away from the $z$ axis of the Bloch sphere, and the measurement errors. In our experiment we extract $\alpha=0.22$ and $\beta=0.42$ from fits to the qubit oscillations. After the probability $P$ is calculated, its maximum should give the value of $\Delta B_z$ during the corresponding record. However, due to the presence of the fluctuating third spin, the probability distribution shows two peaks, at $\Delta B_z^r$ and $\Delta B_z^r + \delta f$, respectively. Their relative weights depend on the time spent by the third spin in its two possible states during that record. As a consequence, the distribution of the differences of $\Delta B_z$ for two consecutive records is a sum of three Gaussian distributions centred on 0 (no change of branch), $\delta f = \SI{1.16}{\mega\hertz}$ (change from $\left| \uparrow \right\rangle \otimes \left|ST\right\rangle$ to $\left| \downarrow \right\rangle \otimes \left|ST\right\rangle$) and $\delta f = \SI{-1.16}{\mega\hertz}$ (change from $\left| \downarrow \right\rangle \otimes \left|ST\right\rangle$ to $\left| \uparrow \right\rangle \otimes \left|ST\right\rangle$) as shown by the black circles in the inset of Fig.~2(b) of the main text. We found from simulations (not shown here) that we can reproduce the observed correlator distribution for transition rates $\Gamma_{\uparrow \rightarrow \downarrow} \sim \SI{1.3}{\kilo\hertz}$ and $\Gamma_{\downarrow \rightarrow \uparrow} \sim \SI{2.6}{\kilo\hertz}$, and a population in the spin-up branch $p_\uparrow \sim 0.7$. The latter is measured from the amplitude of the beating pattern in the qubit oscillations from post-selected records (see Fig.~S\ref{Fig:S1}(b)). This value of 0.7 is consistent with a thermal population for $B_{ext}\sim \SI{0.7}{\tesla}$ and $T_{el}\sim \SI{300}{\milli\kelvin}$ in our experiment.

We find that due to the measurement errors, measurement noise, and system fluctuations, the probability distribution obtained by Eq.~\eqref{eq:Bayes} deviates appreciably from the ideal behaviour just described. The peak assignments according to peak height are not reliable. Fortunately, even though in our measurement the third spin state was not controlled, the fact that it typically flips several times during one record allows us to disentangle its influence on $\Delta B_z$, and identify the value of $\Delta B_{nuc}^z$ we are interested in. To this end, we implemented the following algorithm: 1) We take $N$ consecutive records, for each of which the Bayesian estimator builds up a probability distribution with $k_r$ peaks at positions $f_{k_r}\equiv \Delta B_z(k_r)$ with weights $w_{k_r}$ (taken as the square root of the peak height). 2) We consider all possible combinations $\mathcal{C}$ of assignments of $\Delta B^z_{nuc}$ with the third spin on the lower branch $\left| \uparrow \right\rangle \otimes \left|ST\right\rangle$ to a probability peak in each record. 3) For every $\mathcal{C}$ (``path''), we calculate

\begin{equation*}
\Omega(\mathcal{C})=\prod_{r=1}^{N} \frac{1}{\sqrt{2 \pi \sigma_0^2}} e^{-\frac{\left(f_{k_r(\mathcal{C})} -  f_{k_{r-1}(\mathcal{C})}\right)^2}{2 \sigma_0^2}} w_{k_r(\mathcal{C})}
\end{equation*}
with $\sigma_0=\sigma_c(t_{rec})$ the correlator for one record and $f_{k_{0}}$ the frequency of the record preceding the current sequence, which was estimated in the previous iteration of this algorithm. 4) We identify the path $\mathcal{C}_{max}$ that maximizes $\Omega(\mathcal{C})$ and assign the value of the nuclear field of the first record in the sequence as $\Delta B_z(k_1(\mathcal{C}_{max}))$. 5) We move forward by one index in the register of records and repeat from step 1. We extensively checked this algorithm against simulations and found that for our parameters it performs well with the sequence length of $N=8$, which we used to process the measured data. The resulting nuclear field correlator distribution is shown in the inset of Fig.~2(b) of the main text by green circles, and exhibits a single Gaussian peak as expected. 

\begin{figure}[!]
\includegraphics[width=0.48\textwidth]{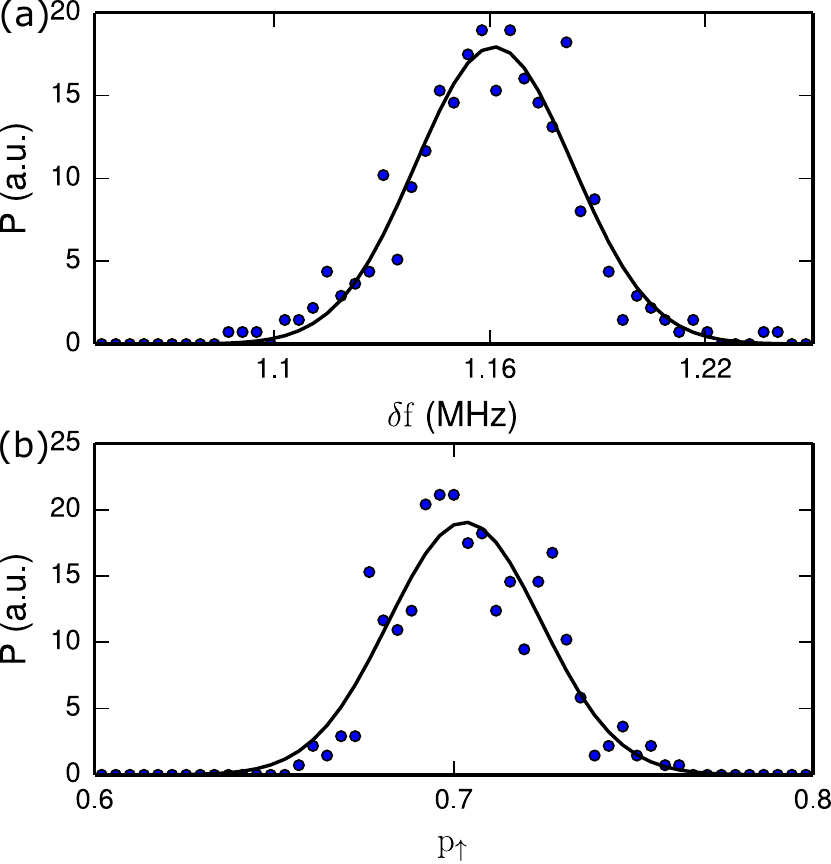}
\caption{Third spin frequency shift and population. (a) Probability density distribution of the qubit frequency difference $\delta f$ between the two branches $\left| \uparrow \right\rangle \otimes \left|ST\right\rangle$ and $\left| \downarrow \right\rangle \otimes \left|ST\right\rangle$. The solid line is a fit to a normal distribution. (b) Probability density distribution of the population $p_{\uparrow}$ of the third spin in the spin-up branch. The solid line is a fit to a normal distribution.} \label{Fig:S1}
\end{figure}

\section{PROBABILITY DISTRIBUTION FOR $T_{2,\phi}^\star$ AND THE RELATED VARIANCE} \label{epaps:Gamma_distribution}
In the data processing, we fit the qubit oscillations decay, which can be parametrized by either $T_{2,\phi}^\star$, or $\sigma_\phi^2$, related by the definition $T_{2,\phi}^\star \equiv 1/\sqrt{2\pi^2 \sigma_{\phi}^2}$. We find by inspection that the probability distribution function of $T_{2,\phi}^\star$ is well described by the Gamma distribution
\begin{equation*}
G_X(x;k,h)=\frac{(2k/h)^{k}}{2^k\Gamma(k)}x^{k-1}e^{-\frac{k}{h}x},
\end{equation*}
where $\Gamma$ is the Euler gamma function, $h$ is the mean, and $k$ is the shape parameter related to the distribution skewness $\gamma_1 = 2/\sqrt{k}$. From here the distribution of $\sigma^2_\phi$ follows as 
\begin{equation*}
\widetilde{G}_Y(y;k,h)=\frac{\alpha}{2}\frac{1}{(\alpha y)^{3/2}} G_X\left(\frac{1}{\sqrt{\alpha y}};k,h\right),
\end{equation*}
where $Y=1/(\alpha X^2)$, with $Y=\sigma_{\phi}^2$, $X=T_{2,\phi}^\star$, and $\alpha = 2 \pi^2$. The fit to the data of both of these functions is shown in Fig.~\ref{Fig:S2}(a) and (b) as red lines. The means of the two distributions are related by 
\begin{equation} \label{eq:Esperance}
\overline{Y}=\frac{1}{ \overline{X} ^2} \frac{k^2}{(k-2)(k-1)},
\end{equation}
which follows by integrating by parts twice. Although $k$ shows small variations with the number of records, we find $k \in \langle 6,8.5 \rangle$ for the whole range of acquisition times considered in the experiment. Using the value in the middle, $k \approx 7.25$,  gives the relation
\begin{equation} \label{eq:Esperance_num}
\overline{\sigma_{\phi}^2} \approx 1.6\frac{1}{2\pi^2 \overline{T_{2,\phi}^\star}^2}.
\end{equation}
We also tried an opposite procedure, fitting $\sigma_{\phi}^2$ to the Gamma distribution and inverting it to get the distribution for $T_{2,\phi}^\star$
\begin{equation*}
\widetilde{G}_Y(y;k,h)=\frac{2}{\alpha y^3} G_X\left(\frac{1}{\alpha y^2};k,h\right)
\end{equation*}
where $Y=1/\sqrt{\alpha X}$ with $Y=T_{2,\phi}^\star$, $X=\sigma_{\phi}^2$ and $\alpha = 2 \pi^2$. The results are plotted as blue lines in Fig.~S\ref{Fig:S2}(a) and (b) and demonstrate a distinctively poorer agreement with the data.

\begin{figure}[!]
\includegraphics[width=0.48\textwidth]{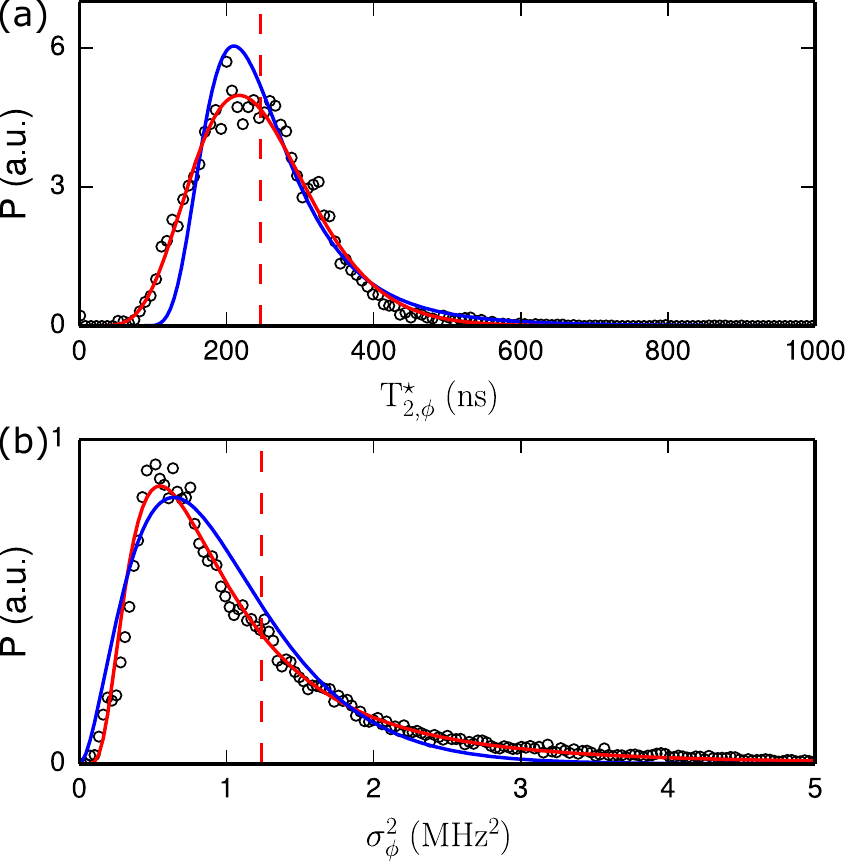}
\caption{Gamma distribution of $T_{2,\phi}^\star$. (a) Distribution of $T_{2,\phi}^\star$ for 100 consecutive records corresponding to $\Delta t=\SI{0.4}{\milli\second}$ as in the lowest panel of Fig.~3(a) of the main text. (b) Distribution of $\sigma_{\phi}^2 = 1/(2\pi^2 {T_{2,\phi}^\star}^2)$ for the same dataset. Red lines correspond to a fit to a Gamma distribution function $G_{T_{2,\phi}^\star}$ in (a) and the corresponding transformed distribution function $\widetilde{G}_{\sigma_{\phi}^2}$ in (b). Blue lines correspond to a fit to a Gamma distribution function $G_{\sigma_{\phi}^2}$ in (b) and the corresponding transformed distribution function $\widetilde{G}_{T_{2,\phi}^\star}$ in (a). The vertical red dashed lines show the mean of both ``red'' distributions, $G_{T_{2,\phi}^\star}$ in (a) and $\widetilde{G}_{\sigma_{\phi}^2}$ in (b).} \label{Fig:S2}
\end{figure}

\section{DERIVATIONS OF THE DEPHASING TIMES RELATIONS} \label{epaps:relations_T2}
\subsection{Relation of the variance of a noise trace to the diffusion speed} \label{epaps:subs:one_over_six}
In the following we need an auxiliary result characterizing a normal diffusion. Let us adopt the standard discrete Gaussian random walk model for the evolution of a variable $\omega$,
\begin{equation} \label{eq:B_chain}
\omega_{n+1} = \omega_n + \delta_n,
\end{equation}
with $\delta_n \sim \mathcal{N}(0, \sigma^2)$ unbiased normally distributed random variables with variance $\sigma^2$. We aim at calculating the variance of the set $\{ \omega_n\}_{n=1}^N$. Using the definition of the variance and the chain defined by Eq.~\eqref{eq:B_chain}, we find
\begin{eqnarray}
\sigma_\omega^2 & = & \frac{1}{N} \sum_{i=1}^{N} \left(\omega_c - \overline{\omega}\right)^2 \nonumber \\*
& = & \sum_{i=1}^{N} \delta_i^2 \kappa_i(1-\kappa_i) + 2 \sum_{i=1}^{N} \sum_{j=1}^{i-1} \delta_i \delta_j \kappa_i(1-\kappa_j), \nonumber \\*
&&
\end{eqnarray}
where $\kappa_i=(N + 1 - i)/N$. Although we were not able to derive the full probability distribution for $\sigma_\omega^2$, its average is straightforward. Using the probability distribution of variables $\delta_n$ we get
\begin{eqnarray} \label{eq:sqrt six}
\overline{\sigma_\omega^2} & = & \sigma^2 \sum_{i=1}^{N} \kappa_i(1-\kappa_i) \nonumber \\* 
& = & \frac{1}{6} \left(1 - \frac{1}{N^2} \right)  N\sigma^2 \nonumber \\*
& \equiv & \frac{1}{6} \left(1 - \frac{1}{N^2} \right) \sigma_N^2,
\end{eqnarray}
where we used the standard result relating the average distance between the final and starting values of $\omega$ in a Gaussian random walk, $\sigma_N^2 = N \sigma^2$.

\subsection{Relation between $T_{2,\phi}^\star$ and the variance of the field correlator $\sigma_B^2$} \label{epaps:subs:T2phi_to_T2B}
We consider $N$ consecutive records containing $M$ measurement points each, using the same indexing as before. The measurement expectation value of the qubit state at time $\tau_c$ is 
\begin{equation} \label{eq:averaging}
s(\tau_c) = \frac{1}{N} \sum_{r=1}^{N} \cos\left(\omega_{c,r} \tau_c\right).
\end{equation}
If $N \gg 1$, the angular frequency values in the set $\{\omega_{c,r} \}_{r=1}^N$ are distributed according to a normal distribution $\mathcal{N}(\overline{\omega_c}, \sigma_c^2)$, fully described by its average and variance,
\begin{eqnarray*}
\overline{\omega_c} & = & \frac{1}{N} \sum_{r=1}^{N} \omega_{c,r}, \\
\sigma_c^2 & = & \frac{1}{N} \sum_{r=1}^{N} \left(\omega_{c,r} -  \overline{\omega_c}\right)^2.
\end{eqnarray*}
Eq.~\eqref{eq:averaging} then gives
\begin{equation} \label{eq:T2black_1}
s(\tau_c)  \approx \cos\left(\overline{\omega_c}\tau_c \right) e^{-\frac{(\tau_c \sigma_c)^2}{2}}.
\end{equation}
In this limit, $\sigma_c^2$ does not practically depend on index $i$, which further gives 
\begin{equation} \label{eq:T2black_2}
s(\tau_c) \approx \cos\left(\overline{\overline{\omega}}\tau_c \right) e^{-\frac{(\tau_c \sigma_c)^2}{2}} e^{-\frac{\tau_c^2}{2}var(\overline{\omega})},
\end{equation}
with the average frequency of the whole trace
\begin{equation*}
\overline{\overline{\omega}} = \frac{1}{M} \sum_{c=1}^{M} \frac{1}{N} \sum_{r=1}^{N} \omega_{c,r}, \\
\end{equation*}
and the variance of the distribution of the $M$ mean frequencies $\overline{\omega_c}$,
\begin{equation*}
var(\overline{\omega}) = \frac{1}{M} \sum_{c=1}^{M} \left( \overline{\omega_c} -  \overline{\overline{\omega}} \right)^2.
\end{equation*}
For large number of records $N$, it holds that $var({\overline{\omega}}) \ll \sigma_c^2$, since these two quantities relate to the fluctuations of the nuclear field within the time of a single record (for which we have no experimental insights) and within the total acquisition time $\Delta t = N t_{rec}$, respectively. This finally gives the relation \begin{equation} \label{eq:sqrt_of_6_theorem}
\overline{\sigma_{\phi}^2}(\Delta t) \equiv \overline{\sigma_c^2+var(\omega)} \approx \overline{\sigma_c^2} \approx  \frac{\sigma_B^2(\Delta t)}{6},
\end{equation}
between the decay of the oscillations (left hand side), and the nuclear field correlator relating the field values displaced by the total acquisition time (right hand side), which we related to $\overline{\sigma_c^2}$ using Eq.~\eqref{eq:sqrt six} for large $N$. 
Combining Eqs.~(\ref{eq:Esperance_num}) and (\ref{eq:sqrt_of_6_theorem}), we can express the previous equation using the dephasing times as
\begin{equation}
\overline{T_{2,\phi}^\star}/T_{2,B}^\star = \sqrt{6} \sqrt{\frac{k^2}{(k-2)(k-1)}} \approx 3.
\end{equation}
This analytical limit value is shown in Fig.~4(b) of the main text as a black dashed line, towards which the data head for large $\Delta t$. The full range for $k \in \langle 6, 8.5 \rangle$ is shown by dotted black lines.

\subsection{Relation between $T_{2,QC}^\star$ and the variance of the field correlator $\sigma_B^2$} \label{epaps:subs:T2QC_to_T2B}
The ensemble of records which defines $T_{2,QC}^\star$ (downwards purple arrows and triangles in Fig.~4) differs from the one defining $T_{2,ps}^\star$ (red arrows) by shifting all the selected chains from the latter ensemble by 1 record.  This removes the initial record of the chain, for which the qubit frequency is known only a posteriori. For the rest of the chain, the frequency is known in advance (up to some uncertainty due to the Bayesian estimator error $\Delta_{est}^2$). This allows, in principle, the implementation of a quantum computation algorithm in a scalable way (unlike the case of post-selection). The measurement result is again described by Eq.~(1) of the main text, however now the statistics of the angular frequencies $\omega_{c,r}$ are very different (and much simpler). Assuming that $\omega$ evolves according to a normal diffusion (with the diffusion constant $D$), each frequency can be considered normal Gaussian variables with average $\overline{\overline{\omega}} = 2\pi f_0$, and variance given by
\begin{equation} \label{eq:sigma_purple}
\sigma^2_{c,r} = (2 \pi)^2 \left( \Delta_{est}^2 + D \left[(c-1)t_r + \xi\tau_c \right] \right),
\end{equation}
with $\xi = t_c / \tau_{250}\approx 4000$ the conversion factor between the laboratory time and the qubit evolution time in our measurement scheme.
The term $D\xi\tau_c$ expresses the fluctuations within a single record which is negligible for $N \gg 1$. Eq.~(1) of the main text then gives
\begin{equation} \label{eq:purple_decay}
s(\tau_c) = \cos(\overline{\overline{\omega}}\tau_c) \frac{1}{N} \sum_{r=1}^{N} e^{-\frac{\tau_c^2 \sigma^2_{c,r}}{2}} \equiv \cos(\overline{\overline{\omega}}\tau_c) f_N(\tau_c).
\end{equation}
We note that the decay envelope $f_N(\tau)$ is not well described by a single Gaussian for any $N >1$, unless the acquisition time is much longer than the decorrelation time. Nevertheless we can still fit it by a Gaussian decay to extract an equivalent $T_2^\star$ and check the consistency of this procedure by, for example, comparing the time $\tau$ at which both functions reach $e^{-1}$. Using the least square fitting amounts to the minimization of the integral
\begin{equation} \label{eq:decay_QC}
I = \int_0^\infty \left[ f_N(\tau) - \exp(-\tau^2\sigma_{fit}^2/2)\right]^2 d\tau,
\end{equation}
with respect to the fit parameter $\sigma_{fit}$. The condition for extremum $\partial I/\partial \sigma_{fit}=0$ gives
\begin{equation} \label{eq:fit_minimization}
\frac{1}{2\sqrt{2}} = \frac{1}{N} \sum_{r=1}^{N} \frac{\sigma_{fit}^3}{(\sigma_{fit}^2 +\sigma^2_{c,r})^{3/2}}.
\end{equation}
Using  Eq.~(\ref{eq:sigma_purple}), and transforming the sum over $r$ to an integral, appropriate for $N\gg1$, we find
\begin{eqnarray}
\frac{T_{2,QC}^\star}{T_{2,B}^\star} & \equiv & \sqrt{\frac{\sigma_B^2}{\sigma_{fit}^2}} \nonumber \\*
& = & \sqrt{\frac{1}{2} \left(8 \sqrt{2} - 1 - \sqrt{1 + 16\sqrt{2}} \right)} \nonumber \\*
& \approx & 1.65.
\end{eqnarray}
Fig.~4(b) shows the agreement between this value (dashed purple line) and the data for a large number of records. To extend the correspondence regime of validity, we perform a numerical evaluation of Eq.~(\ref{eq:fit_minimization}). The result is shown by a solid purple line which agrees very well with our data for any $N$. We want to point out that although we didn't access it, all the dephasing times discussed here should merge into a single value $T_{2,\infty}^\star$ in the ergodic regime, at long times.

\section{DETERMINATION OF $T_2^\star$ IN POST-SELECTION} \label{epaps:obtaining_T2ps}

\begin{figure}[!]
\includegraphics[width=0.48\textwidth]{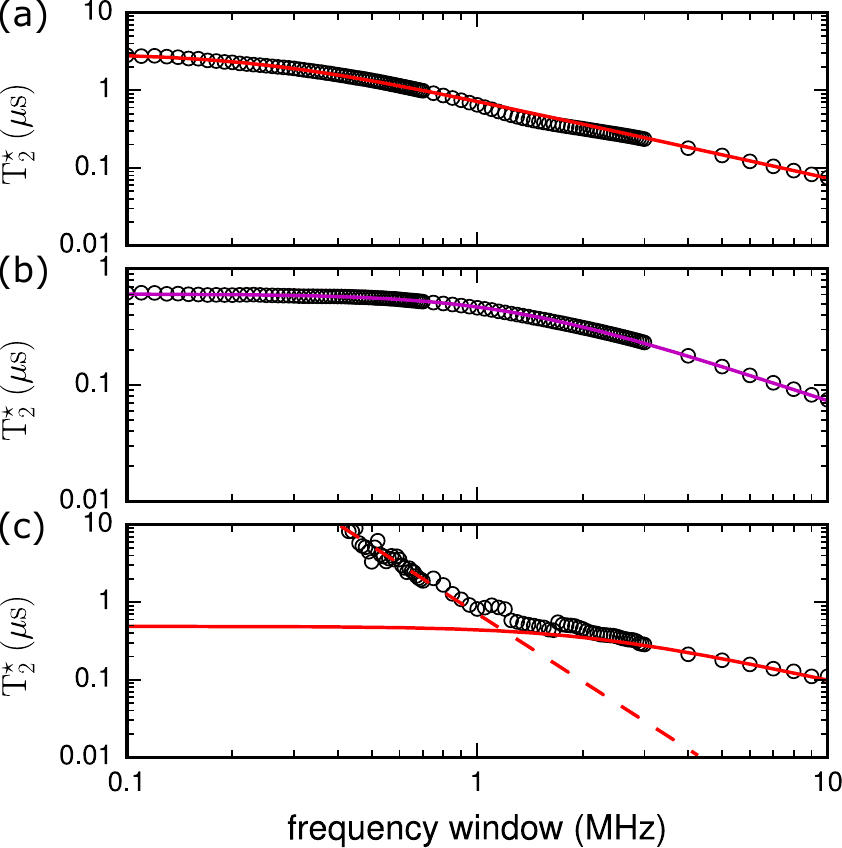}
\caption{Determination of $T_2^\star$ from post-selection. Open circles are $T_2^\star$ extracted from decaying oscillations as shown in the lower insets of Fig.~4 of the main text by varying the frequency window $\Delta f$ in which the records are selected. (a), Case of the averaging of the post-selected records giving $T_{2,ps}^\star$ (upwards red triangles in Fig.~4(a) of the main text), (b), Case of the averaging of the subsequent records giving $T_{2,QC}^\star$ (downwards purple triangles in Fig.~4(a) of the main text). (c), Same as (a) with a constraint on the Bayesian estimator probability distribution peak below $\SI{200}{kHz}$ as shown in Fig.~4(c) of the main text. Solid lines are a fit to Eq.~(\ref{eq:Windowing}). The dashed line in (c) is a fit to a power law with exponent $T_2^\star \propto \Delta f^{-3}$.} \label{Fig:S3}
\end{figure}

When selecting records at a particular frequency $f_0$ within a frequency window $\Delta f$, the latter plays the role of an artificial broadening of the field fluctuations. The decay of oscillations is therefore due to the contributions of the fluctuations of the field, characterized by $\sigma_\phi^2(\Delta t)$ with the acquisition time $\Delta t=N t_{rec}$ of $N$ records, the Bayesian estimator error $\Delta_{est}$, and the frequency window $\Delta f$. We therefore write
\begin{equation} \label{eq:Windowing}
T_2^\star(\Delta t,\Delta f)= \frac{1}{\sqrt{2\pi^2}\sqrt{\sigma_\phi^2(\Delta t) + \Delta_{est}^2 + \gamma \Delta f^2}}.
\end{equation}
We fit this formula as a function of $\Delta f$ with $\gamma$ and $\sigma_\phi^2(\Delta t) + \Delta_{est}^2$ as fitting parameters, and define $T_{2,ps}^\star$ as the value fitted at $\Delta f \to 0$. The coefficient $\gamma$ arises because as $\Delta f$ is increased, more and more records are inserted into the post-selection ensemble. It quickly results in an ensemble of chains of consecutive records, rather than individual (statistically independent) records. For a chain of records, the relation between the decay and the frequency variance contains the non-trivial coefficient $\sqrt{6}$ from Eq.~(\ref{eq:sqrt_of_6_theorem}), which is the reason for $\gamma\neq 1$ here. Combining Eqs.~(\ref{eq:Esperance_num}) and (\ref{eq:sqrt_of_6_theorem}) we get
\begin{equation}
\gamma = \frac{1}{6} \frac{(k-2)(k-1)}{k^2}.
\end{equation}
As $k \in \langle 6, 8.5\rangle$, we expect $\gamma \in \langle 0.09, 0.11 \rangle$, in excellent agreement with the value $\gamma \approx 0.09$ that we obtain from the fit to the data (see Fig.~\ref{Fig:S3}(a) and (b)).

Apart from a different rule for the choice of records for the selected ensemble, $T_{2,QC}^\star$ is defined by the same fitting procedure as just described. On the other hand, when adding the constraint on the width of the Bayesian estimator probability peak, records are progressively removed from the ensemble of selected records as the constraint becomes stricter. In this case, increasing $\Delta f$ does not result in consecutive chains of records, and Eq.~(\ref{eq:Windowing}) breaks down as shown in Fig.~\ref{Fig:S3}(c). Interestingly, in this regime $T_2^\star$ exhibits a power law dependence, $T_2^\star \propto \Delta f^{-3}$. Finally, when $T_2^\star$ gets too large compared to the maximal evolution time $\tau_{250}$, the decay becomes too small to extract a reliable value. We therefore define $T_{2,ps}^\star$ as the last value that falls on the power law. This is how we determine the value of $T_{2,ps}^\star \geq \SI{10}{\micro\second}$ reported in the main text, as a conservative estimate for the longest $T_{2,ps}^\star$.

\end{widetext}


\begin{thebibliography}{36}%
\makeatletter
\providecommand \@ifxundefined [1]{%
 \@ifx{#1\undefined}
}%
\providecommand \@ifnum [1]{%
 \ifnum #1\expandafter \@firstoftwo
 \else \expandafter \@secondoftwo
 \fi
}%
\providecommand \@ifx [1]{%
 \ifx #1\expandafter \@firstoftwo
 \else \expandafter \@secondoftwo
 \fi
}%
\providecommand \natexlab [1]{#1}%
\providecommand \enquote  [1]{``#1''}%
\providecommand \bibnamefont  [1]{#1}%
\providecommand \bibfnamefont [1]{#1}%
\providecommand \citenamefont [1]{#1}%
\providecommand \href@noop [0]{\@secondoftwo}%
\providecommand \href [0]{\begingroup \@sanitize@url \@href}%
\providecommand \@href[1]{\@@startlink{#1}\@@href}%
\providecommand \@@href[1]{\endgroup#1\@@endlink}%
\providecommand \@sanitize@url [0]{\catcode `\\12\catcode `\$12\catcode
  `\&12\catcode `\#12\catcode `\^12\catcode `\_12\catcode `\%12\relax}%
\providecommand \@@startlink[1]{}%
\providecommand \@@endlink[0]{}%
\providecommand \url  [0]{\begingroup\@sanitize@url \@url }%
\providecommand \@url [1]{\endgroup\@href {#1}{\urlprefix }}%
\providecommand \urlprefix  [0]{URL }%
\providecommand \Eprint [0]{\href }%
\providecommand \doibase [0]{http://dx.doi.org/}%
\providecommand \selectlanguage [0]{\@gobble}%
\providecommand \bibinfo  [0]{\@secondoftwo}%
\providecommand \bibfield  [0]{\@secondoftwo}%
\providecommand \translation [1]{[#1]}%
\providecommand \BibitemOpen [0]{}%
\providecommand \bibitemStop [0]{}%
\providecommand \bibitemNoStop [0]{.\EOS\space}%
\providecommand \EOS [0]{\spacefactor3000\relax}%
\providecommand \BibitemShut  [1]{\csname bibitem#1\endcsname}%
\let\auto@bib@innerbib\@empty
\bibitem [{\citenamefont {Abragam}(1961)}]{Abragam}%
  \BibitemOpen
  \bibfield  {author} {\bibinfo {author} {\bibfnamefont {A.}~\bibnamefont
  {Abragam}},\ }\href@noop {} {\emph {\bibinfo {title} {{The Principles of
  Nuclear Magnetism}}}}\ (\bibinfo  {publisher} {Oxford University Press},\
  \bibinfo {year} {1961})\ p.\ \bibinfo {pages} {618}\BibitemShut {NoStop}%
\bibitem [{\citenamefont {Poole}(1993)}]{Poole1993}%
  \BibitemOpen
  \bibfield  {author} {\bibinfo {author} {\bibfnamefont {C.}~\bibnamefont
  {Poole}},\ }\href@noop {} {\emph {\bibinfo {title} {{Electron Spin
  Resonance}}}},\ \bibinfo {edition} {2nd}\ ed.\ (\bibinfo  {publisher} {Wiley,
  New York},\ \bibinfo {year} {1993})\BibitemShut {NoStop}%
\bibitem [{\citenamefont {DiVincenzo}(1995)}]{DiVincenzo1995}%
  \BibitemOpen
  \bibfield  {author} {\bibinfo {author} {\bibfnamefont {D.~P.}\ \bibnamefont
  {DiVincenzo}},\ }\href {\doibase 10.1126/science.270.5234.255} {\bibfield
  {journal} {\bibinfo  {journal} {Science}\ }\textbf {\bibinfo {volume}
  {270}},\ \bibinfo {pages} {255} (\bibinfo {year} {1995})}\BibitemShut
  {NoStop}%
\bibitem [{\citenamefont {Loss}\ and\ \citenamefont
  {DiVincenzo}(1998)}]{Loss1998}%
  \BibitemOpen
  \bibfield  {author} {\bibinfo {author} {\bibfnamefont {D.}~\bibnamefont
  {Loss}}\ and\ \bibinfo {author} {\bibfnamefont {D.~P.}\ \bibnamefont
  {DiVincenzo}},\ }\href {\doibase 10.1103/PhysRevA.57.120} {\bibfield
  {journal} {\bibinfo  {journal} {Phys. Rev. A}\ }\textbf {\bibinfo {volume}
  {57}},\ \bibinfo {pages} {120} (\bibinfo {year} {1998})}\BibitemShut
  {NoStop}%
\bibitem [{\citenamefont {Petta}\ \emph {et~al.}(2005)\citenamefont {Petta},
  \citenamefont {Johnson}, \citenamefont {Taylor}, \citenamefont {Laird},
  \citenamefont {Yacoby}, \citenamefont {Lukin}, \citenamefont {Marcus},
  \citenamefont {Hanson},\ and\ \citenamefont {Gossard}}]{Petta2005}%
  \BibitemOpen
  \bibfield  {author} {\bibinfo {author} {\bibfnamefont {J.~R.}\ \bibnamefont
  {Petta}}, \bibinfo {author} {\bibfnamefont {A.~C.}\ \bibnamefont {Johnson}},
  \bibinfo {author} {\bibfnamefont {J.~M.}\ \bibnamefont {Taylor}}, \bibinfo
  {author} {\bibfnamefont {E.~A.}\ \bibnamefont {Laird}}, \bibinfo {author}
  {\bibfnamefont {A.}~\bibnamefont {Yacoby}}, \bibinfo {author} {\bibfnamefont
  {M.~D.}\ \bibnamefont {Lukin}}, \bibinfo {author} {\bibfnamefont {C.~M.}\
  \bibnamefont {Marcus}}, \bibinfo {author} {\bibfnamefont {M.~P.}\
  \bibnamefont {Hanson}}, \ and\ \bibinfo {author} {\bibfnamefont {A.~C.}\
  \bibnamefont {Gossard}},\ }\href {\doibase 10.1126/science.1116955}
  {\bibfield  {journal} {\bibinfo  {journal} {Science}\ }\textbf {\bibinfo
  {volume} {309}},\ \bibinfo {pages} {2180} (\bibinfo {year}
  {2005})}\BibitemShut {NoStop}%
\bibitem [{\citenamefont {Merkulov}\ \emph {et~al.}(2002)\citenamefont
  {Merkulov}, \citenamefont {Efros},\ and\ \citenamefont
  {Rosen}}]{Merkulov2002}%
  \BibitemOpen
  \bibfield  {author} {\bibinfo {author} {\bibfnamefont {I.~A.}\ \bibnamefont
  {Merkulov}}, \bibinfo {author} {\bibfnamefont {A.~L.}\ \bibnamefont {Efros}},
  \ and\ \bibinfo {author} {\bibfnamefont {M.}~\bibnamefont {Rosen}},\ }\href
  {\doibase 10.1103/PhysRevB.65.205309} {\bibfield  {journal} {\bibinfo
  {journal} {Phys. Rev. B}\ }\textbf {\bibinfo {volume} {65}},\ \bibinfo
  {pages} {205309} (\bibinfo {year} {2002})}\BibitemShut {NoStop}%
\bibitem [{\citenamefont {Khaetskii}\ \emph {et~al.}(2002)\citenamefont
  {Khaetskii}, \citenamefont {Loss},\ and\ \citenamefont
  {Glazman}}]{Khaetskii2002}%
  \BibitemOpen
  \bibfield  {author} {\bibinfo {author} {\bibfnamefont {A.~V.}\ \bibnamefont
  {Khaetskii}}, \bibinfo {author} {\bibfnamefont {D.}~\bibnamefont {Loss}}, \
  and\ \bibinfo {author} {\bibfnamefont {L.}~\bibnamefont {Glazman}},\ }\href
  {\doibase 10.1103/PhysRevLett.88.186802} {\bibfield  {journal} {\bibinfo
  {journal} {Phys. Rev. Lett.}\ }\textbf {\bibinfo {volume} {88}},\ \bibinfo
  {pages} {186802} (\bibinfo {year} {2002})}\BibitemShut {NoStop}%
\bibitem [{\citenamefont {Urbaszek}\ \emph {et~al.}(2013)\citenamefont
  {Urbaszek}, \citenamefont {Marie}, \citenamefont {Amand}, \citenamefont
  {Krebs}, \citenamefont {Voisin}, \citenamefont {Maletinsky}, \citenamefont
  {H\"{o}gele},\ and\ \citenamefont {Imamoglu}}]{Urbaszek2013}%
  \BibitemOpen
  \bibfield  {author} {\bibinfo {author} {\bibfnamefont {B.}~\bibnamefont
  {Urbaszek}}, \bibinfo {author} {\bibfnamefont {X.}~\bibnamefont {Marie}},
  \bibinfo {author} {\bibfnamefont {T.}~\bibnamefont {Amand}}, \bibinfo
  {author} {\bibfnamefont {O.}~\bibnamefont {Krebs}}, \bibinfo {author}
  {\bibfnamefont {P.}~\bibnamefont {Voisin}}, \bibinfo {author} {\bibfnamefont
  {P.}~\bibnamefont {Maletinsky}}, \bibinfo {author} {\bibfnamefont
  {A.}~\bibnamefont {H\"{o}gele}}, \ and\ \bibinfo {author} {\bibfnamefont
  {A.}~\bibnamefont {Imamoglu}},\ }\href {\doibase 10.1103/RevModPhys.85.79}
  {\bibfield  {journal} {\bibinfo  {journal} {Rev. Mod. Phys.}\ }\textbf
  {\bibinfo {volume} {85}},\ \bibinfo {pages} {79} (\bibinfo {year}
  {2013})}\BibitemShut {NoStop}%
\bibitem [{\citenamefont {Paget}\ \emph {et~al.}(1977)\citenamefont {Paget},
  \citenamefont {Lampel}, \citenamefont {Sapoval},\ and\ \citenamefont
  {Safarov}}]{Paget1977}%
  \BibitemOpen
  \bibfield  {author} {\bibinfo {author} {\bibfnamefont {D.}~\bibnamefont
  {Paget}}, \bibinfo {author} {\bibfnamefont {G.}~\bibnamefont {Lampel}},
  \bibinfo {author} {\bibfnamefont {B.}~\bibnamefont {Sapoval}}, \ and\
  \bibinfo {author} {\bibfnamefont {V.~I.}\ \bibnamefont {Safarov}},\ }\href
  {\doibase 10.1103/PhysRevB.15.5780} {\bibfield  {journal} {\bibinfo
  {journal} {Phys. Rev. B}\ }\textbf {\bibinfo {volume} {15}},\ \bibinfo
  {pages} {5780} (\bibinfo {year} {1977})}\BibitemShut {NoStop}%
\bibitem [{\citenamefont {Maletinsky}\ \emph {et~al.}(2007)\citenamefont
  {Maletinsky}, \citenamefont {Badolato},\ and\ \citenamefont
  {Imamoglu}}]{Maletinsky2007a}%
  \BibitemOpen
  \bibfield  {author} {\bibinfo {author} {\bibfnamefont {P.}~\bibnamefont
  {Maletinsky}}, \bibinfo {author} {\bibfnamefont {A.}~\bibnamefont
  {Badolato}}, \ and\ \bibinfo {author} {\bibfnamefont {A.}~\bibnamefont
  {Imamoglu}},\ }\href {\doibase 10.1103/PhysRevLett.99.056804} {\bibfield
  {journal} {\bibinfo  {journal} {Phys. Rev. Lett.}\ }\textbf {\bibinfo
  {volume} {99}},\ \bibinfo {pages} {056804} (\bibinfo {year}
  {2007})}\BibitemShut {NoStop}%
\bibitem [{\citenamefont {Bluhm}\ \emph {et~al.}(2011)\citenamefont {Bluhm},
  \citenamefont {Foletti}, \citenamefont {Neder}, \citenamefont {Rudner},
  \citenamefont {Mahalu}, \citenamefont {Umansky},\ and\ \citenamefont
  {Yacoby}}]{Bluhm2010}%
  \BibitemOpen
  \bibfield  {author} {\bibinfo {author} {\bibfnamefont {H.}~\bibnamefont
  {Bluhm}}, \bibinfo {author} {\bibfnamefont {S.}~\bibnamefont {Foletti}},
  \bibinfo {author} {\bibfnamefont {I.}~\bibnamefont {Neder}}, \bibinfo
  {author} {\bibfnamefont {M.}~\bibnamefont {Rudner}}, \bibinfo {author}
  {\bibfnamefont {D.}~\bibnamefont {Mahalu}}, \bibinfo {author} {\bibfnamefont
  {V.}~\bibnamefont {Umansky}}, \ and\ \bibinfo {author} {\bibfnamefont
  {A.}~\bibnamefont {Yacoby}},\ }\href {\doibase 10.1038/nphys1856} {\bibfield
  {journal} {\bibinfo  {journal} {Nat. Phys.}\ }\textbf {\bibinfo {volume}
  {7}},\ \bibinfo {pages} {109} (\bibinfo {year} {2011})}\BibitemShut {NoStop}%
\bibitem [{\citenamefont {de~Sousa}(2009)}]{DeSousa2009}%
  \BibitemOpen
  \bibfield  {author} {\bibinfo {author} {\bibfnamefont {R.}~\bibnamefont
  {de~Sousa}},\ }\href {\doibase 10.1007/978-3-540-79365-6\_10} {\bibfield
  {journal} {\bibinfo  {journal} {Top. Appl. Phys.}\ }\textbf {\bibinfo
  {volume} {115}},\ \bibinfo {pages} {183} (\bibinfo {year}
  {2009})}\BibitemShut {NoStop}%
\bibitem [{\citenamefont {Shulman}\ \emph {et~al.}(2014)\citenamefont
  {Shulman}, \citenamefont {Harvey}, \citenamefont {Nichol}, \citenamefont
  {Bartlett}, \citenamefont {Doherty}, \citenamefont {Umansky},\ and\
  \citenamefont {Yacoby}}]{Shulman2014}%
  \BibitemOpen
  \bibfield  {author} {\bibinfo {author} {\bibfnamefont {M.~D.}\ \bibnamefont
  {Shulman}}, \bibinfo {author} {\bibfnamefont {S.~P.}\ \bibnamefont {Harvey}},
  \bibinfo {author} {\bibfnamefont {J.~M.}\ \bibnamefont {Nichol}}, \bibinfo
  {author} {\bibfnamefont {S.~D.}\ \bibnamefont {Bartlett}}, \bibinfo {author}
  {\bibfnamefont {A.~C.}\ \bibnamefont {Doherty}}, \bibinfo {author}
  {\bibfnamefont {V.}~\bibnamefont {Umansky}}, \ and\ \bibinfo {author}
  {\bibfnamefont {A.}~\bibnamefont {Yacoby}},\ }\href {\doibase
  10.1038/ncomms6156} {\bibfield  {journal} {\bibinfo  {journal} {Nat.
  Commun.}\ }\textbf {\bibinfo {volume} {5}},\ \bibinfo {pages} {5156}
  (\bibinfo {year} {2014})}\BibitemShut {NoStop}%
\bibitem [{\citenamefont {Pioro-Ladri\`{e}re}\ \emph
  {et~al.}(2008)\citenamefont {Pioro-Ladri\`{e}re}, \citenamefont {Obata},
  \citenamefont {Tokura}, \citenamefont {Shin}, \citenamefont {Kubo},
  \citenamefont {Yoshida}, \citenamefont {Taniyama},\ and\ \citenamefont
  {Tarucha}}]{Pioro-Ladriere2008}%
  \BibitemOpen
  \bibfield  {author} {\bibinfo {author} {\bibfnamefont {M.}~\bibnamefont
  {Pioro-Ladri\`{e}re}}, \bibinfo {author} {\bibfnamefont {T.}~\bibnamefont
  {Obata}}, \bibinfo {author} {\bibfnamefont {Y.}~\bibnamefont {Tokura}},
  \bibinfo {author} {\bibfnamefont {Y.-S.}\ \bibnamefont {Shin}},
  \bibinfo {author} {\bibfnamefont {T.}~\bibnamefont {Kubo}}, \bibinfo {author}
  {\bibfnamefont {K.}~\bibnamefont {Yoshida}}, \bibinfo {author} {\bibfnamefont
  {T.}~\bibnamefont {Taniyama}}, \ and\ \bibinfo {author} {\bibfnamefont
  {S.}~\bibnamefont {Tarucha}},\ }\href {\doibase 10.1038/nphys1053} {\bibfield
   {journal} {\bibinfo  {journal} {Nat. Phys.}\ }\textbf {\bibinfo {volume}
  {4}},\ \bibinfo {pages} {776} (\bibinfo {year} {2008})}\BibitemShut {NoStop}%
\bibitem [{\citenamefont {Hanson}\ \emph {et~al.}(2007)\citenamefont {Hanson},
  \citenamefont {Kouwenhoven}, \citenamefont {Petta}, \citenamefont {Tarucha},\
  and\ \citenamefont {Vandersypen}}]{Hanson2007}%
  \BibitemOpen
  \bibfield  {author} {\bibinfo {author} {\bibfnamefont {R.}~\bibnamefont
  {Hanson}}, \bibinfo {author} {\bibfnamefont {L.~P.}\ \bibnamefont
  {Kouwenhoven}}, \bibinfo {author} {\bibfnamefont {J.~R.}\ \bibnamefont
  {Petta}}, \bibinfo {author} {\bibfnamefont {S.}~\bibnamefont {Tarucha}}, \
  and\ \bibinfo {author} {\bibfnamefont {L.~M.~K.}\ \bibnamefont
  {Vandersypen}},\ }\href {\doibase 10.1103/RevModPhys.79.1217} {\bibfield
  {journal} {\bibinfo  {journal} {Rev. Mod. Phys.}\ }\textbf {\bibinfo {volume}
  {79}},\ \bibinfo {pages} {1217} (\bibinfo {year} {2007})}\BibitemShut
  {NoStop}%
\bibitem [{\citenamefont {Barthel}\ \emph {et~al.}(2012)\citenamefont
  {Barthel}, \citenamefont {Medford}, \citenamefont {Bluhm}, \citenamefont
  {Yacoby}, \citenamefont {Marcus}, \citenamefont {Hanson},\ and\ \citenamefont
  {Gossard}}]{Barthel2012}%
  \BibitemOpen
  \bibfield  {author} {\bibinfo {author} {\bibfnamefont {C.}~\bibnamefont
  {Barthel}}, \bibinfo {author} {\bibfnamefont {J.}~\bibnamefont {Medford}},
  \bibinfo {author} {\bibfnamefont {H.}~\bibnamefont {Bluhm}}, \bibinfo
  {author} {\bibfnamefont {A.}~\bibnamefont {Yacoby}}, \bibinfo {author}
  {\bibfnamefont {C.~M.}\ \bibnamefont {Marcus}}, \bibinfo {author}
  {\bibfnamefont {M.~P.}\ \bibnamefont {Hanson}}, \ and\ \bibinfo {author}
  {\bibfnamefont {A.~C.}\ \bibnamefont {Gossard}},\ }\href {\doibase
  10.1103/PhysRevB.85.035306} {\bibfield  {journal} {\bibinfo  {journal} {Phys.
  Rev. B}\ }\textbf {\bibinfo {volume} {85}},\ \bibinfo {pages} {035306}
  (\bibinfo {year} {2012})}\BibitemShut {NoStop}%
\bibitem [{\citenamefont {Sergeevich}\ \emph {et~al.}(2011)\citenamefont
  {Sergeevich}, \citenamefont {Chandran}, \citenamefont {Combes}, \citenamefont
  {Bartlett},\ and\ \citenamefont {Wiseman}}]{Sergeevich2011}%
  \BibitemOpen
  \bibfield  {author} {\bibinfo {author} {\bibfnamefont {A.}~\bibnamefont
  {Sergeevich}}, \bibinfo {author} {\bibfnamefont {A.}~\bibnamefont
  {Chandran}}, \bibinfo {author} {\bibfnamefont {J.}~\bibnamefont {Combes}},
  \bibinfo {author} {\bibfnamefont {S.~D.}\ \bibnamefont {Bartlett}}, \ and\
  \bibinfo {author} {\bibfnamefont {H.~M.}\ \bibnamefont {Wiseman}},\ }\href
  {\doibase 10.1103/PhysRevA.84.052315} {\bibfield  {journal} {\bibinfo
  {journal} {Phys. Rev. A}\ }\textbf {\bibinfo {volume} {84}},\ \bibinfo
  {pages} {052315} (\bibinfo {year} {2011})}\BibitemShut {NoStop}%
\bibitem [{\citenamefont {Wang}\ and\ \citenamefont
  {Uhlenbeck}(1945)}]{Wang1945}%
  \BibitemOpen
  \bibfield  {author} {\bibinfo {author} {\bibfnamefont {M.~C.}\ \bibnamefont
  {Wang}}\ and\ \bibinfo {author} {\bibfnamefont {G.~E.}\ \bibnamefont
  {Uhlenbeck}},\ }\href {\doibase 10.1103/RevModPhys.17.323} {\bibfield
  {journal} {\bibinfo  {journal} {Rev. Mod. Phys.}\ }\textbf {\bibinfo {volume}
  {17}},\ \bibinfo {pages} {323} (\bibinfo {year} {1945})}\BibitemShut
  {NoStop}%
\bibitem [{\citenamefont {Redfield}(1959)}]{Redfield1959}%
  \BibitemOpen
  \bibfield  {author} {\bibinfo {author} {\bibfnamefont {A.~G.}\ \bibnamefont
  {Redfield}},\ }\href {\doibase 10.1103/PhysRev.116.315} {\bibfield  {journal}
  {\bibinfo  {journal} {Phys. Rev.}\ }\textbf {\bibinfo {volume} {116}},\
  \bibinfo {pages} {315} (\bibinfo {year} {1959})}\BibitemShut {NoStop}%
\bibitem [{\citenamefont {Klauser}\ \emph {et~al.}(2008)\citenamefont
  {Klauser}, \citenamefont {Coish},\ and\ \citenamefont {Loss}}]{Klauser2008}%
  \BibitemOpen
  \bibfield  {author} {\bibinfo {author} {\bibfnamefont {D.}~\bibnamefont
  {Klauser}}, \bibinfo {author} {\bibfnamefont {W.~A.}\ \bibnamefont {Coish}},
  \ and\ \bibinfo {author} {\bibfnamefont {D.}~\bibnamefont {Loss}},\ }\href
  {\doibase 10.1103/PhysRevB.78.205301} {\bibfield  {journal} {\bibinfo
  {journal} {Phys. Rev. B}\ }\textbf {\bibinfo {volume} {78}},\ \bibinfo
  {pages} {205301} (\bibinfo {year} {2008})}\BibitemShut {NoStop}%
\bibitem [{\citenamefont {Bouchaud}\ and\ \citenamefont
  {Georges}(1990)}]{Bouchaud1990}%
  \BibitemOpen
  \bibfield  {author} {\bibinfo {author} {\bibfnamefont {J.-P.}\ \bibnamefont
  {Bouchaud}}\ and\ \bibinfo {author} {\bibfnamefont {A.}~\bibnamefont
  {Georges}},\ }\href {\doibase 10.1016/0370-1573(90)90099-N} {\bibfield
  {journal} {\bibinfo  {journal} {Phys. Rep.}\ }\textbf {\bibinfo {volume}
  {195}},\ \bibinfo {pages} {127} (\bibinfo {year} {1990})}\BibitemShut
  {NoStop}%
\bibitem [{\citenamefont {Metzler}\ and\ \citenamefont
  {Klafter}(2000)}]{Metzler2000}%
  \BibitemOpen
  \bibfield  {author} {\bibinfo {author} {\bibfnamefont {R.}~\bibnamefont
  {Metzler}}\ and\ \bibinfo {author} {\bibfnamefont {J.}~\bibnamefont
  {Klafter}},\ }\href {\doibase 10.1016/S0370-1573(00)00070-3} {\bibfield
  {journal} {\bibinfo  {journal} {Phys. Rep.}\ }\textbf {\bibinfo {volume}
  {339}},\ \bibinfo {pages} {1} (\bibinfo {year} {2000})}\BibitemShut {NoStop}%
\bibitem [{\citenamefont {Li}\ \emph {et~al.}(2012)\citenamefont {Li},
  \citenamefont {Sinitsyn}, \citenamefont {Smith}, \citenamefont {Reuter},
  \citenamefont {Wieck}, \citenamefont {Yakovlev}, \citenamefont {Bayer},\ and\
  \citenamefont {Crooker}}]{Li2012}%
  \BibitemOpen
  \bibfield  {author} {\bibinfo {author} {\bibfnamefont {Y.}~\bibnamefont
  {Li}}, \bibinfo {author} {\bibfnamefont {N.}~\bibnamefont {Sinitsyn}},
  \bibinfo {author} {\bibfnamefont {D.~L.}\ \bibnamefont {Smith}}, \bibinfo
  {author} {\bibfnamefont {D.}~\bibnamefont {Reuter}}, \bibinfo {author}
  {\bibfnamefont {A.~D.}\ \bibnamefont {Wieck}}, \bibinfo {author}
  {\bibfnamefont {D.~R.}\ \bibnamefont {Yakovlev}}, \bibinfo {author}
  {\bibfnamefont {M.}~\bibnamefont {Bayer}}, \ and\ \bibinfo {author}
  {\bibfnamefont {S.~A.}\ \bibnamefont {Crooker}},\ }\href {\doibase
  10.1103/PhysRevLett.108.186603} {\bibfield  {journal} {\bibinfo  {journal}
  {Phys. Rev. Lett.}\ }\textbf {\bibinfo {volume} {108}},\ \bibinfo {pages}
  {186603} (\bibinfo {year} {2012})}\BibitemShut {NoStop}%
\bibitem [{\citenamefont {Reilly}\ \emph {et~al.}(2008)\citenamefont {Reilly},
  \citenamefont {Taylor}, \citenamefont {Laird}, \citenamefont {Petta},
  \citenamefont {Marcus}, \citenamefont {Hanson},\ and\ \citenamefont
  {Gossard}}]{Reilly2008}%
  \BibitemOpen
  \bibfield  {author} {\bibinfo {author} {\bibfnamefont {D.~J.}\ \bibnamefont
  {Reilly}}, \bibinfo {author} {\bibfnamefont {J.~M.}\ \bibnamefont {Taylor}},
  \bibinfo {author} {\bibfnamefont {E.~A.}\ \bibnamefont {Laird}}, \bibinfo
  {author} {\bibfnamefont {J.~R.}\ \bibnamefont {Petta}}, \bibinfo {author}
  {\bibfnamefont {C.~M.}\ \bibnamefont {Marcus}}, \bibinfo {author}
  {\bibfnamefont {M.~P.}\ \bibnamefont {Hanson}}, \ and\ \bibinfo {author}
  {\bibfnamefont {A.~C.}\ \bibnamefont {Gossard}},\ }\href {\doibase
  10.1103/PhysRevLett.101.236803} {\bibfield  {journal} {\bibinfo  {journal}
  {Phys. Rev. Lett.}\ }\textbf {\bibinfo {volume} {101}},\ \bibinfo {pages}
  {236803} (\bibinfo {year} {2008})}\BibitemShut {NoStop}%
\bibitem [{\citenamefont {Medford}\ \emph {et~al.}(2012)\citenamefont
  {Medford}, \citenamefont {Cywiński}, \citenamefont {Barthel}, \citenamefont
  {Marcus}, \citenamefont {Hanson},\ and\ \citenamefont
  {Gossard}}]{Medford2012}%
  \BibitemOpen
  \bibfield  {author} {\bibinfo {author} {\bibfnamefont {J.}~\bibnamefont
  {Medford}}, \bibinfo {author} {\bibfnamefont {A.}~\bibnamefont {Cywiński}},
  \bibinfo {author} {\bibfnamefont {C.}~\bibnamefont {Barthel}}, \bibinfo
  {author} {\bibfnamefont {C.~M.}\ \bibnamefont {Marcus}}, \bibinfo {author}
  {\bibfnamefont {M.~P.}\ \bibnamefont {Hanson}}, \ and\ \bibinfo {author}
  {\bibfnamefont {A.~C.}\ \bibnamefont {Gossard}},\ }\href {\doibase
  10.1103/PhysRevLett.108.086802} {\bibfield  {journal} {\bibinfo  {journal}
  {Phys. Rev. Lett.}\ }\textbf {\bibinfo {volume} {108}},\ \bibinfo {pages}
  {086802} (\bibinfo {year} {2012})}\BibitemShut {NoStop}%
\bibitem [{EPA()}]{EPAPS}%
  \BibitemOpen
  \href@noop {} {\ }\bibinfo {note} {See Supplemental Material}\BibitemShut
  {NoStop}%
\bibitem [{\citenamefont {Hahn}(1950)}]{Hahn1950}%
  \BibitemOpen
  \bibfield  {author} {\bibinfo {author} {\bibfnamefont {E.}~\bibnamefont
  {Hahn}},\ }\href {\doibase 10.1103/PhysRev.80.580} {\bibfield  {journal}
  {\bibinfo  {journal} {Phys. Rev.}\ }\textbf {\bibinfo {volume} {80}},\
  \bibinfo {pages} {580} (\bibinfo {year} {1950})}\BibitemShut {NoStop}%
\bibitem [{\citenamefont {Vandersypen}\ and\ \citenamefont
  {Chuang}(2005)}]{Vandersypen2005}%
  \BibitemOpen
  \bibfield  {author} {\bibinfo {author} {\bibfnamefont {L.~M.~K.}\
  \bibnamefont {Vandersypen}}\ and\ \bibinfo {author} {\bibfnamefont {I.~L.}\
  \bibnamefont {Chuang}},\ }\href {\doibase 10.1103/RevModPhys.76.1037}
  {\bibfield  {journal} {\bibinfo  {journal} {Rev. Mod. Phys.}\ }\textbf
  {\bibinfo {volume} {76}},\ \bibinfo {pages} {1037} (\bibinfo {year}
  {2005})}\BibitemShut {NoStop}%
\bibitem [{\citenamefont {Maune}\ \emph {et~al.}(2012)\citenamefont {Maune},
  \citenamefont {Borselli}, \citenamefont {Huang}, \citenamefont {Ladd},
  \citenamefont {Deelman}, \citenamefont {Holabird}, \citenamefont {Kiselev},
  \citenamefont {Alvarado-Rodriguez}, \citenamefont {Ross}, \citenamefont
  {Schmitz}, \citenamefont {Sokolich}, \citenamefont {Watson}, \citenamefont
  {Gyure},\ and\ \citenamefont {Hunter}}]{Maune2012}%
  \BibitemOpen
  \bibfield  {author} {\bibinfo {author} {\bibfnamefont {B.~M.}\ \bibnamefont
  {Maune}}, \bibinfo {author} {\bibfnamefont {M.~G.}\ \bibnamefont {Borselli}},
  \bibinfo {author} {\bibfnamefont {B.}~\bibnamefont {Huang}}, \bibinfo
  {author} {\bibfnamefont {T.~D.}\ \bibnamefont {Ladd}}, \bibinfo {author}
  {\bibfnamefont {P.~W.}\ \bibnamefont {Deelman}}, \bibinfo {author}
  {\bibfnamefont {K.~S.}\ \bibnamefont {Holabird}}, \bibinfo {author}
  {\bibfnamefont {A.~A.}\ \bibnamefont {Kiselev}}, \bibinfo {author}
  {\bibfnamefont {I.}~\bibnamefont {Alvarado-Rodriguez}}, \bibinfo {author}
  {\bibfnamefont {R.~S.}\ \bibnamefont {Ross}}, \bibinfo {author}
  {\bibfnamefont {A.~E.}\ \bibnamefont {Schmitz}}, \bibinfo {author}
  {\bibfnamefont {M.}~\bibnamefont {Sokolich}}, \bibinfo {author}
  {\bibfnamefont {C.~A.}\ \bibnamefont {Watson}}, \bibinfo {author}
  {\bibfnamefont {M.~F.}\ \bibnamefont {Gyure}}, \ and\ \bibinfo {author}
  {\bibfnamefont {A.~T.}\ \bibnamefont {Hunter}},\ }\href {\doibase
  10.1038/nature10707} {\bibfield  {journal} {\bibinfo  {journal} {Nature}\
  }\textbf {\bibinfo {volume} {481}},\ \bibinfo {pages} {344} (\bibinfo {year}
  {2012})}\BibitemShut {NoStop}%
\bibitem [{\citenamefont {Kawakami}\ \emph {et~al.}(2014)\citenamefont
  {Kawakami}, \citenamefont {Scarlino}, \citenamefont {Ward}, \citenamefont
  {Braakman}, \citenamefont {Savage}, \citenamefont {Lagally}, \citenamefont
  {Friesen}, \citenamefont {Coppersmith}, \citenamefont {Eriksson},\ and\
  \citenamefont {Vandersypen}}]{Kawakami2014}%
  \BibitemOpen
  \bibfield  {author} {\bibinfo {author} {\bibfnamefont {E.}~\bibnamefont
  {Kawakami}}, \bibinfo {author} {\bibfnamefont {P.}~\bibnamefont {Scarlino}},
  \bibinfo {author} {\bibfnamefont {D.~R.}\ \bibnamefont {Ward}}, \bibinfo
  {author} {\bibfnamefont {F.~R.}\ \bibnamefont {Braakman}}, \bibinfo {author}
  {\bibfnamefont {D.~E.}\ \bibnamefont {Savage}}, \bibinfo {author}
  {\bibfnamefont {M.~G.}\ \bibnamefont {Lagally}}, \bibinfo {author}
  {\bibfnamefont {M.}~\bibnamefont {Friesen}}, \bibinfo {author} {\bibfnamefont
  {S.~N.}\ \bibnamefont {Coppersmith}}, \bibinfo {author} {\bibfnamefont
  {M.~A.}\ \bibnamefont {Eriksson}}, \ and\ \bibinfo {author} {\bibfnamefont
  {L.~M.~K.}\ \bibnamefont {Vandersypen}},\ }\href {\doibase
  10.1038/nnano.2014.153} {\bibfield  {journal} {\bibinfo  {journal} {Nat.
  Nanotechnol.}\ }\textbf {\bibinfo {volume} {9}},\ \bibinfo {pages} {666}
  (\bibinfo {year} {2014})}\BibitemShut {NoStop}%
\bibitem [{\citenamefont {Veldhorst}\ \emph {et~al.}(2014)\citenamefont
  {Veldhorst}, \citenamefont {Hwang}, \citenamefont {Yang}, \citenamefont
  {Leenstra}, \citenamefont {de~Ronde}, \citenamefont {Dehollain},
  \citenamefont {Muhonen}, \citenamefont {Hudson}, \citenamefont {Itoh},
  \citenamefont {Morello},\ and\ \citenamefont {Dzurak}}]{Veldhorst2014}%
  \BibitemOpen
  \bibfield  {author} {\bibinfo {author} {\bibfnamefont {M.}~\bibnamefont
  {Veldhorst}}, \bibinfo {author} {\bibfnamefont {J.~C.~C.}\ \bibnamefont
  {Hwang}}, \bibinfo {author} {\bibfnamefont {C.~H.}\ \bibnamefont {Yang}},
  \bibinfo {author} {\bibfnamefont {A.~W.}\ \bibnamefont {Leenstra}}, \bibinfo
  {author} {\bibfnamefont {B.}~\bibnamefont {de~Ronde}}, \bibinfo {author}
  {\bibfnamefont {J.~P.}\ \bibnamefont {Dehollain}}, \bibinfo {author}
  {\bibfnamefont {J.~T.}\ \bibnamefont {Muhonen}}, \bibinfo {author}
  {\bibfnamefont {F.~E.}\ \bibnamefont {Hudson}}, \bibinfo {author}
  {\bibfnamefont {K.~M.}\ \bibnamefont {Itoh}}, \bibinfo {author}
  {\bibfnamefont {A.}~\bibnamefont {Morello}}, \ and\ \bibinfo {author}
  {\bibfnamefont {A.~S.}\ \bibnamefont {Dzurak}},\ }\href {\doibase
  10.1038/nnano.2014.216} {\bibfield  {journal} {\bibinfo  {journal} {Nat.
  Nanotechnol.}\ }\textbf {\bibinfo {volume} {9}},\ \bibinfo {pages} {981}
  (\bibinfo {year} {2014})}\BibitemShut {NoStop}%
\bibitem [{\citenamefont {Delbecq}\ \emph {et~al.}(2014)\citenamefont
  {Delbecq}, \citenamefont {Nakajima}, \citenamefont {Otsuka}, \citenamefont
  {Amaha}, \citenamefont {Watson}, \citenamefont {Manfra},\ and\ \citenamefont
  {Tarucha}}]{Delbecq2014}%
  \BibitemOpen
  \bibfield  {author} {\bibinfo {author} {\bibfnamefont {M.~R.}\ \bibnamefont
  {Delbecq}}, \bibinfo {author} {\bibfnamefont {T.}~\bibnamefont {Nakajima}},
  \bibinfo {author} {\bibfnamefont {T.}~\bibnamefont {Otsuka}}, \bibinfo
  {author} {\bibfnamefont {S.}~\bibnamefont {Amaha}}, \bibinfo {author}
  {\bibfnamefont {J.~D.}\ \bibnamefont {Watson}}, \bibinfo {author}
  {\bibfnamefont {M.~J.}\ \bibnamefont {Manfra}}, \ and\ \bibinfo {author}
  {\bibfnamefont {S.}~\bibnamefont {Tarucha}},\ }\href {\doibase
  10.1063/1.4875909} {\bibfield  {journal} {\bibinfo  {journal} {Appl. Phys.
  Lett.}\ }\textbf {\bibinfo {volume} {104}},\ \bibinfo {pages} {183111}
  (\bibinfo {year} {2014})}\BibitemShut {NoStop}%
\bibitem [{\citenamefont {Otsuka}\ \emph {et~al.}(2015)\citenamefont {Otsuka},
  \citenamefont {Nakajima}, \citenamefont {Delbecq}, \citenamefont {Amaha},
  \citenamefont {Yoneda}, \citenamefont {Takeda}, \citenamefont {Allison},
  \citenamefont {Ito}, \citenamefont {Sugawara}, \citenamefont {Noiri},
  \citenamefont {Ludwig}, \citenamefont {Wieck},\ and\ \citenamefont
  {Tarucha}}]{Otsuka2015a}%
  \BibitemOpen
  \bibfield  {author} {\bibinfo {author} {\bibfnamefont {T.}~\bibnamefont
  {Otsuka}}, \bibinfo {author} {\bibfnamefont {T.}~\bibnamefont {Nakajima}},
  \bibinfo {author} {\bibfnamefont {M.~R.}\ \bibnamefont {Delbecq}}, \bibinfo
  {author} {\bibfnamefont {S.}~\bibnamefont {Amaha}}, \bibinfo {author}
  {\bibfnamefont {J.}~\bibnamefont {Yoneda}}, \bibinfo {author} {\bibfnamefont
  {K.}~\bibnamefont {Takeda}}, \bibinfo {author} {\bibfnamefont
  {G.}~\bibnamefont {Allison}}, \bibinfo {author} {\bibfnamefont
  {T.}~\bibnamefont {Ito}}, \bibinfo {author} {\bibfnamefont {R.}~\bibnamefont
  {Sugawara}}, \bibinfo {author} {\bibfnamefont {A.}~\bibnamefont {Noiri}},
  \bibinfo {author} {\bibfnamefont {A.}~\bibnamefont {Ludwig}}, \bibinfo
  {author} {\bibfnamefont {A.~D.}\ \bibnamefont {Wieck}}, \ and\ \bibinfo
  {author} {\bibfnamefont {S.}~\bibnamefont {Tarucha}},\ }\href
  {http://arxiv.org/abs/1510.02547} {\ ,\ \bibinfo {pages} {1} (\bibinfo {year}
  {2015})},\ \Eprint {http://arxiv.org/abs/1510.02547} {arXiv:1510.02547}
  \BibitemShut {NoStop}%
\bibitem [{\citenamefont {Yao}\ \emph {et~al.}(2006)\citenamefont {Yao},
  \citenamefont {Liu},\ and\ \citenamefont {Sham}}]{Yao2006}%
  \BibitemOpen
  \bibfield  {author} {\bibinfo {author} {\bibfnamefont {W.}~\bibnamefont
  {Yao}}, \bibinfo {author} {\bibfnamefont {R.-B.}\ \bibnamefont {Liu}}, \ and\
  \bibinfo {author} {\bibfnamefont {L.~J.}\ \bibnamefont {Sham}},\ }\href
  {\doibase 10.1103/PhysRevB.74.195301} {\bibfield  {journal} {\bibinfo
  {journal} {Phys. Rev. B}\ }\textbf {\bibinfo {volume} {74}},\ \bibinfo
  {pages} {195301} (\bibinfo {year} {2006})}\BibitemShut {NoStop}%
\bibitem [{\citenamefont {Coish}\ and\ \citenamefont {Loss}(2004)}]{Coish2004}%
  \BibitemOpen
  \bibfield  {author} {\bibinfo {author} {\bibfnamefont {W.~A.}\ \bibnamefont
  {Coish}}\ and\ \bibinfo {author} {\bibfnamefont {D.}~\bibnamefont {Loss}},\
  }\href {\doibase 10.1103/PhysRevB.70.195340} {\bibfield  {journal} {\bibinfo
  {journal} {Phys. Rev. B}\ }\textbf {\bibinfo {volume} {70}},\ \bibinfo
  {pages} {195340} (\bibinfo {year} {2004})}\BibitemShut {NoStop}%
\bibitem [{\citenamefont {Deng}\ and\ \citenamefont {Hu}(2006)}]{Deng2006}%
  \BibitemOpen
  \bibfield  {author} {\bibinfo {author} {\bibfnamefont {C.}~\bibnamefont
  {Deng}}\ and\ \bibinfo {author} {\bibfnamefont {X.}~\bibnamefont {Hu}},\
  }\href {\doibase 10.1103/PhysRevB.73.241303} {\bibfield  {journal} {\bibinfo
  {journal} {Phys. Rev. B}\ }\textbf {\bibinfo {volume} {73}},\ \bibinfo
  {pages} {241303} (\bibinfo {year} {2006})}\BibitemShut {NoStop}%
\end{thebibliography}
\end{document}